\documentclass[%
reprint,
superscriptaddress,
nofootinbib,
amsmath,
amssymb,
aps,
floatfix,
showkeys,
]{revtex4-2}

\usepackage[colorlinks,citecolor=blue,linkcolor=blue,anchorcolor=blue,filecolor=blue,
urlcolor=blue]{hyperref}

\usepackage{graphicx,color,xcolor}
\usepackage{amsfonts,amsmath}
\usepackage{amssymb,ulem}
\usepackage{dcolumn}
\usepackage{bm,lipsum}
\usepackage[utf8]{inputenc}
\usepackage{subeqnarray}
\usepackage{array}
\usepackage{epsfig,hyperref}
\usepackage{morefloats}
\usepackage{relsize}
\usepackage{url}

\usepackage{comment}

\begin{document}

\preprint{ }

\title{Role of local anisotropy in hybrid stars }

\author{Luiz L. Lopes}
\email{llopes@cefetmg.br}
\affiliation{Centro Federal de Educa\c{c}\~ao Tecnol\'ogica de Minas Gerais Campus VIII, Varginha/MG, CEP 37.022-560, Brazil}

\author{H. C. Das}
\email{harishchandra.das@ct.infn.it}
\affiliation{INFN Sezione di Catania, Dipartimento di Fisica, Via S. Sofia 64, 95123 Catania, Italy}
\date{\today}
\begin{abstract}
Using the Bower-Liang model, we discuss how pressure anisotropies affect the microscopic and macroscopic properties of hybrid stars. We find that anisotropies affect the maximum mass, central density, and radius of the canonical stars.  Anisotropies also affect the minimum neutron star mass that presents quarks in their core, as well as the total amount of quarks for the maximally massive stars. We also confront our results with standard constraints, such as the radius and the tidal parameter of the canonical star, as well as the mass and radius of the PSR J0740+6620 pulsar. We observe that moderate values for anisotropies could fulfill these constraints simultaneously. On the other hand, within more extreme degrees of anisotropies,  more speculative constraints such as black widow pulsars PSR J0952-0607 and the mass-gap object in the GW190814 event can be explained as hybrid stars. We also investigate the role of anisotropies in the neutron stars' moment of inertia.
\end{abstract}
\maketitle
\maketitle
\section{Introduction}
Neutron stars are compact and degenerate objects whose density at their core can reach several times the density of the atomic nucleus. The theory of the birth of neutron stars is well-established in the literature as supernova remnants. However, other aspects are far from being determined. One of them is the neutron stars' internal composition. The standard model considers neutron stars composed of nucleons and leptons with zero total charge net and in chemical equilibrium. Still, the presence of exotic, non-Nucleonic degrees of freedom is possible to exist in neutron stars' core. One of the vastly studied possibilities is the presence of hyperons~\cite{PAND1971,Glen1991,Schaffner_1996,WeissenbornNPA_2012,lopes2023PRDa}. Indeed, some studies suggest that hyperons are inevitable~\cite{Apo_2010,LOPES2021NPA}.

Another possibility is that at high densities, the quark matter becomes energetically favorable, creating a hadron-quark phase transition in the neutron star's core. In this case, we have a neutron star constituted by deconfined quark matter in the core, surrounded by hadronic matter in the outer layers. Such a configuration is called a hybrid star~\cite{GREINER1998NPA,Glen1992PRD,Lopes_2022}.
The study of hybrid stars is essential; a recent study pointed out that the presence of deconfined quarks in massive neutron stars is not only possible but probable~\cite{Annala_2020}.

Another unknown feature of neutron stars is related to their isotropic pressure. Most works (see ref.~\cite{Glenbook}) assume that neutron stars are isotropic objects. However, several exotic phenomena, such as a strong magnetic field, superfluity, phase transitions, and others (see ref.~\cite{Herrera_1997} for an extensive discussion), can be present in neutron stars' interiors, inducing a local anisotropy. { Several models have been proposed to explore the effects of pressure anisotropy inside the star \cite{Das_ILC_2022, Horowitz_2014, Cosenza_1981}. One of them is the so-called Bower-Liang (BL) model~\cite{Bowers_1974}. In the BL model, the anisotropy is local and gravitationally induced, meaning that it arises due to the variations in the gravitational field within the star, which affect the local pressure distribution ~\cite{DelgadoeDelgado2018}. While various phenomena, as mentioned above, can introduce local anisotropy and break spherical symmetry, however, the BL model specifically considers anisotropy that does not compromise the overall spherical symmetry of the star, although this introduces complexities when reconciling with Pascal's law.}
 
In this work, we study the effects of local anisotropies in different configurations of hybrid stars. In the hadronic phase, we construct hybrid stars with and without hyperons, while in the quark phase, we use one soft and one stiff equation of state (EOS). To accomplish this task, we use the quantum hadrodynamic (QHD) for the hadron phase~\cite{Serot_1992}, and the thermodynamic consistent vector MIT bag model for the quark phase~\cite{Lopes_2021, Lopes_2021b}. The presence of the anisotropy is taken into account via BL model ~\cite{Bowers_1974}.

All our results are confronted with experimental and observational constraints. From the microscopic nuclear physics point of view, there are four well-constrained parameters at the saturation density~\cite{Oertel2017}: the saturation density itself ($n_0$), the binding energy per nucleon ($E/A)$, the compressibility ($K$) and the effective nucleon mass ($M^*/M$). Two additional parameters, the symmetry energy $(S_0)$ and its slope $(L)$, are still a matter of debate. Nevertheless, they are strongly constrained in a recent study~\cite{Essick2021}.

From the macroscopic nuclear astrophysics point of view, there are also some constraints. Maybe the more important is the  PSR J0740+6620 with a mass of $2.08 \pm 0.07 \  M_\odot$ and a radius of $12.35 \pm 0.75$ km ~\cite{Miller_2021}. Concerning the canonical $1.4 \ M_\odot$ star, two NICER results point that their radius must be in the range $11.52 <R_{1.4}< 13.85$ km~\cite{Riley_2019}, and $11.96 <R_{1.4}< 14.26$ km~\cite{Miller_2019}. Later, these results were refined to $11.80 <R_{1.4}< 13.10$ km ~\cite{Miller_2021}. Besides these standard constraints, there are some unorthodox and speculative ones presented in the literature. The first is the HESS J1731-347 object~\cite{Doroshenko_2022} whose mass and radius are $M=0.77_{-0.17}^{+0.20}~M_\odot$ and $R = 10.4 _{-0.78}^{+0.86}$ km respectively. The true nature of the HESS object is one of the hot topics in nuclear astrophysics. In ref.~\cite{Doroshenko_2022}, it was shown that some chiral theory can explain this object as an ordinary hadronic neutron star. The hadronic nature was also explored in ref.~\cite{Kubis_2023}. On the other hand, the HESS object as a strange star was studied in refs.~\cite{Lopes2023M, Rather2023, Lopes2024EPJC}. Finally, in ref.~\cite{Sagun2023} the authors show that the
HESS object can be a hadronic neutron star with a soft EOS or a hybrid star with an early deconfinement phase transition.  A second one is the speculative mass of the black widow pulsars PSR J0952-0607, $M = 2.35\pm0.17 M_\odot$ \cite{Romani_2022}.  Finally, there is the mass-gap object in the GW190814 event~\cite{RAbbott_2020}, whose mass was estimated to be $2.50 - 2.67 \ M_\odot$. Although we are not sure about its true nature~\cite{Lopes_ApJ, Das_PRD_2021}. We check if some degree of anisotropy can explain some or even all of these exotic objects as hybrid stars.

In addition, we also use the constraints from the gravitational wave observations by LIGO/VIRGO/KAGRA. The GW170817 event~\cite{Abbott_2017} put the constraints on the dimensionless tidal parameter of the canonical star  $\Lambda_{1.4}<800$~\cite{Abbott_2017}. This result was then refined in ref.~\cite{Abbott_2018}, to  $70<\Lambda_{1.4} <580$. Moreover, assuming that the mass-gap object in the GW190814 event was not a black hole implies that the dimensionless tidal parameter for the canonical star must be in the range of $458 <\Lambda_{1.4}<889$~\cite{RAbbott_2020}. 

Ultimately, we discuss the moment of inertia (MOI) of the hybrid stars. Till now, we don't have any observational data for any NS. The authors of ref.~\cite{Landry_2018} have obtained the MOI for several pulsars using the universal relations between the mass and the tidal deformability. Here, we calculate the MOI of the anisotropic hybrid stars with different hadronic and quark parametrizations by varying the degrees of anisotropicity. One can constrain the value of MOI from the future observational data for the anisotropic compact star. There are other ways to constrain the magnitude of the MOI for different systems, such as Millisecond Pulsars (MSP), Double NS (DNS), and Low Mass X-ray Binary (LMXB), as done in Refs.~\cite{Landry_2018, Kumar_2019}. The MOI of these pulsars is expected to be measured soon; for example, the PSR J0737-3039(A) is the only known DNS up to date. Since the NS EOS is believed to be universal, the tidal deformability constraints from GW170817 have implications for all NSs, including PSR J0737-3039(A), the MOI has been obtained to be $1.15_{-0.24}^{+0.38}\times 10^{45}$ g cm$^2$ \cite{Landry_2018}. One can also estimate the MOI for other MSPs and LMXBs.
\section{Microscopic Formalism}
\subsection{Quantum Hadrodynamics}
To simulate the interaction between baryons in dense cold matter, we use an extended version of the QHD model, whose Lagrangian density reads~\citep{Serot_1992, Fattoyev_2010}:
\begin{align}
\mathcal{L}_{\rm QHD} &= \bar{\psi}_B[\gamma^\mu(i\partial_\mu  - g_{B\omega}\omega_\mu   - g_{B\rho} \frac{1}{2}\vec{\tau} \cdot \vec{\rho}_\mu)
\nonumber \\
&- (M_B - g_{B\sigma}\sigma)]\psi_B  -U(\sigma)   \nonumber   \\
&+ \frac{1}{2}(\partial_\mu \sigma \partial^\mu \sigma - m_s^2\sigma^2) - \frac{1}{4}\Omega^{\mu \nu}\Omega_{\mu \nu} + \frac{1}{2} m_v^2 \omega_\mu \omega^\mu 
\nonumber \\
&+ \frac{1}{2} m_\rho^2 \vec{\rho}_\mu \cdot \vec{\rho}^{ \; \mu} - \frac{1}{4}\bf{P}^{\mu \nu} \cdot \bf{P}_{\mu \nu} + \mathcal{L}_{\omega\rho} + \mathcal{L}_{\phi} , \label{s1} 
\end{align}
in natural units. $\psi_B$  is the baryonic  Dirac field, where $B$ can stand either for nucleons only ($N$) or can run over $N$ and hyperons ($H$). The $\sigma$, $\omega_\mu$ and $\vec{\rho}_\mu$ are the mesonic fields, while $\vec{\tau}$ are the Pauli matrices. The $g's$ are the Yukawa coupling constants that simulate the strong interaction, $M_B$ is the baryon mass,  $m_s$, $m_v$, and $m_\rho$ are the masses of the $\sigma$, $\omega$ and $\rho$ mesons respectively. The $U(\sigma)$ is the self-interaction term introduced in \citet{Boguta_1977}:
\begin{eqnarray}
U(\sigma) =  \frac{1}{3}\lambda \sigma^3 + \frac{1}{4}\kappa \sigma^4, \nonumber 
\end{eqnarray}
and $\mathcal{L}_{\omega\rho}$ is a non-linear $\omega$-$\rho$ coupling interaction as in ref.~\cite{Fattoyev_2010}:
\begin{eqnarray}
 \mathcal{L}_{\omega\rho} = \Lambda_{\omega\rho}(g_{N\rho}^2 \vec{\rho^\mu} \cdot \vec{\rho_\mu}) (g_{N\omega}^2 \omega^\mu \omega_\mu) , \label{EL2}
\end{eqnarray}
which is necessary to correct the slope of the symmetry energy and has a strong influence on the radii and tidal deformation of the neutron stars~\citep{Cavagnoli2011, Dexheimer_2019, Lopes2023PRDb}; $\mathcal{L}_\phi$ is related to the strangeness of hidden $\phi$ vector meson, which couples only with the hyperons, not affecting the properties of symmetric nuclear matter:
\begin{equation}
\mathcal{L}_\phi = g_{H \phi}\bar{\psi}_H(\gamma^\mu\phi_\mu)\psi_H + \frac{1}{2}m_\phi^2\phi_\mu\phi^\mu - \frac{1}{4}\Phi^{\mu\nu}\Phi_{\mu\nu} , \label{EL3} 
\end{equation}
as pointed out in refs. ~\cite{Lopes2020EPJA, WeissenbornNPA_2012}, this vector channel is crucial to obtain massive Hyperonic neutron stars. 

As neutron stars are stable macroscopic objects, we need to describe a neutral, chemically stable matter, and hence, leptons are added as free Fermi gases, whose Lagrangian density is the usual one.
\begin{eqnarray}
 \mathcal{L}_l =  \bar{\psi}_l[\gamma^\mu(i\partial_\mu  - m_l)]\psi_l. \nonumber 
\end{eqnarray}

To solve the equations of motion, we use the mean field approximation (MFA), where the meson fields are replaced by their expectation values. Applying the Euler-Lagrange formalism and using the quantization rules ($E = \partial^0$, $k = i\partial^j$), we easily obtain the eigenvalue for the energy:
\begin{equation}
 E_B = \sqrt{k^2 + M_B^{*2}} + g_{B\omega}\omega_0 + g_{B\phi}\phi_0 + \frac{\tau_{3B}}{2}g_{B\rho}\rho_0 , \label{EL4}
\end{equation}
where $M^{*}_B~\equiv~M_B - g_{B\sigma}\sigma_0$ is the effective baryon mass and $\tau_{3B}$ assumes the value of +1 for p, $\Sigma^+$, $\Sigma^0$  and $\Xi^0$;  and -1 for n, $\Lambda^0$ $\Sigma^{-}$  and $\Xi^-$. Within this approach, we can add the mixed term $g_{\Sigma\Lambda\rho}$ as discussed in ref.~\cite{Lopes2023PTEP}. This term is necessary to recover the relations of completeness and closure.
For the leptons, we have:
\begin{equation}
 E_l =  \sqrt{k^2+m_l^2}, \label{EL5}
\end{equation}
and the mesonic fields in MFA are given by:
\begin{align}
&m_\sigma^2\sigma_0 + \lambda\sigma_0^2 +\kappa\sigma_0^3 = \sum_B g_{B\sigma}n_B^s , \nonumber \\
&(m_\omega^2 + 2\Lambda_v\rho_0^2)\omega = \sum_B g_{B\omega}n_B ,
\nonumber \\
&m_\phi^2\phi_0  = \sum_B g_{B \phi} n_B,  \nonumber \\
&(m_\rho^2 +2\Lambda_v\omega_0^2)\rho_0 = \sum g_{B\rho}\frac{\tau_{3B}}{2}n_B, \label{EL6}
\end{align}
where $\Lambda_v~\equiv~\Lambda_{\omega\rho}g_{N\omega}^2g_{N\rho}^2$, and $n_B^s$ and $n_B$ are, respectively, the scalar density and the number density of the baryon $B$. Finally, applying Fermi-Dirac
statistics to baryons and leptons and with the help of Eq.~(\ref{EL6}), we can write the total energy density as~\citep{Miyatsu2013}:
\begin{align}
\epsilon &= \sum_B \frac{1}{\pi^2}\int_0^{k_{Bf}} dk k^2 \sqrt{k^2 + M_B^{*2}} +U(\sigma_0)
\nonumber \\
&+\frac{1}{2}m_\sigma^2\sigma_0^2 + \frac{1}{2}m_\omega^2\omega_0^2 + \frac{1}{2}m_\phi^2\phi_0^2 + \frac{1}{2}m_\rho^2\rho_0^2
\nonumber \\
&+3 \Lambda_v\omega_0^2\rho_0^2 
\label{EL7}  + \sum_l \frac{1}{\pi^2}\int_0^{k_{lf}} dk k^2 \sqrt{k^2 + m_l^{2}} , 
\end{align}
and the pressure is easily obtained by thermodynamic relations: $p =
\sum_f \mu_f n_f - \epsilon$, where the sum runs over all the fermions, and $\mu_f$ is the corresponding chemical potential. Now, to determine each particle population, we impose that the matter is $\beta$ stable and has a total electric net charge equal to zero.
\subsection{Parametrization}
In this work, we use the enhanced L3$\omega\rho$ parametrization (eL3$\omega\rho$), as originally proposed in ref.~\cite{lopes2023PRDa}, such parametrization virtually fulfills all constraints at the saturation density. The parameters, constraints, and predictions of this model are presented in Tab.~\ref{TL0}.
\begin{table}[ht]
\centering
\begin{tabular}{c|c}
\hline
\multicolumn{2}{c}{Enhanced L3$\omega\rho$}
\\ \hline
$(g_{N\sigma}/m_s)^2$ & 12.108 fm$^2$ \\
$(g_{N\omega}/m_v)^2$ & 7.132  fm$^2$ \\
$(g_{N\rho}/m_\rho)^2$ & 5.85  fm$^2$ \\
$\kappa$ & -25.919 \\
$\lambda$ &  14.501 fm$^{-1}$ \\
$\Lambda_{\omega\rho}$ &  0.0283 \\ 
\end{tabular}
\begin{tabular}{c|cc}
\hline 
Quantity & Constraint & This model \\ \hline
$n_0$ ($fm^{-3}$) & 0.148--0.170 & 0.156 \\
$M^{*}/M$ & 0.60--0.80 & 0.69  \\
$K$ (MeV)& 220--260 &  256  \\
$S_0$ (MeV) & 31.2--35.0 &  32.1  \\
$L$ (MeV) & 38--67 & 66\\
$B/A$ (MeV) & 15.8--16.5  & 16.2  \\ 
\hline
\end{tabular}
\caption{eL3$\omega\rho$~\cite{lopes2023PRDa} parameterization (top) and its predictions for nuclear matter quantities (bottom). The phenomenological constraints are taken from Ref.~\cite{Oertel2017, Essick2021}.} 
\label{TL0}
\end{table}

If hyperons are present, we must fix their coupling constants with the mesonic fields as well. The coupling constants of the vector mesons can be fixed within the flavor SU(3) group symmetry, { although the flavor SU(3) symmetry is broken at mass level, once the members of the baryon octet present different masses,  the calculation of the coupling constants do not depend of the masses values}. Assuming an ideal angle mixing, $\theta_v = 35.264$, and a rating $z = 1\sqrt{6}$  the only free parameter left is the $\alpha_v$. As we wish to produce neutron stars with $M\geq 2.0 \ M_\odot$, we use $\alpha_v$ = 0.50 { (the definition and how to fix $z$, $\theta_V$ and $\alpha_V$ are out the scope of the present work. We refer the interested reader to see refs.~\cite{Dover_1984,Weissenborn_2012, Miyatsu2013,LOPES2021NPA,lopes2023PRDa,Lopes2023PTEP} and the references therein to it, as well  for a more complete discussion about group theory and its relation with the meson-hyperon coupling constants)}.

For the coupling with the scalar meson, the hyperon potential depths are taken as $U_\Lambda = -28$ MeV, $U_\Sigma = +30$ MeV~\cite{Schaffner_2000}, and $U_\Xi = -4$ MeV~\cite{Inoue2019} respectively. The complete set of coupling constants used in this work are given as follows:
\begin{align}
&g_{\Lambda\omega}/g_{N\omega} = 0.714, \, g_{\Sigma\omega}/g_{N\omega} = 1.0, \, g_{\Xi\omega}/g_{N\omega} = 0.571,
\nonumber \\
&g_{\Lambda\rho} = 0.0, \quad g_{\Sigma\rho} = 1.0, \quad g_{\Xi\rho}/g_{N\rho} = 0.0,
\nonumber \\
&g_{\Lambda\sigma}/g_{N\sigma} = 0.646, \, g_{\Sigma\sigma}/g_{N\sigma} = 0.663, \, g_{\Xi\sigma}/g_{N\sigma} = 0.453,
\nonumber \\
&g_{\Lambda\phi}/g_{N\omega} = -0.808, \, g_{\Sigma\phi}/g_{N\omega} = -0.404,
\nonumber \\
&g_{\Xi\phi}/g_{N\omega} = -1.01, \, g_{\Sigma\Lambda\rho} = 0.577, 
\end{align}
where we explicitly add the mixed Yukawa coupling in order to restore the relations of completeness and closure of the SU(3) Clebsch-Gordan (CG) coefficients, { i.e, the sum of the squared CG for a given interaction must be one, $\sum C_i^2 = 1$, where the $C's$ are the CG coefficients. A complete discussion about this feature can be found in ref.~\cite{Lopes2023PTEP} and the references therein.}

\subsection{Vector MIT bag model}
The quark matter is studied within the vector MIT bag model, which is an extension of the original MIT bag model~\cite{MITbag} that incorporates some
features of the QHD. In its original form, the MIT bag model considers that each baryon is composed of three non-interacting quarks inside a bag. The
bag, in turn, corresponds to an infinite potential that confines the quarks. As a consequence, the quarks are free inside the bag and are forbidden to reach its exterior. All the information about the strong force relies on the bag pressure value, which mimics the vacuum pressure.

In the vector MIT bag model,  { an effective model,} the quarks are still confined inside the bag, but now they interact with each other through a vector meson exchange. This vector meson plays a role analog to the $\omega$ meson of the QHD~\cite{Serot_1992}. Moreover, the contribution of the Dirac sea can be taken into account through a self-interaction of the vector meson~\cite{furnstahl1997vacuum}. The Lagrangian of the vector MIT bag model, therefore, consists of the Lagrangian of the original MIT, plus the Yukawa-type Lagrangian of the vector field exchange, plus the Dirac sea contribution. We must also add the mesonic mass term to maintain the thermodynamic consistency. It then reads~\cite{Lopes_2021,Lopes_2021b}:
\begin{equation}
\mathcal{L} =  \mathcal{L}_{\rm MIT} +  \mathcal{L}_{\rm V} + \mathcal{L}_{\rm DIRAC}, \label{l1}
\end{equation}
where
\begin{equation}
\mathcal{L}_{\rm MIT} = \sum_{i}\{ \bar{\psi}_i  [ i\gamma^{\mu} \partial_\mu - m_i ]\psi_i - B \}\Theta(\bar{\psi_i}\psi_i), \label{e1}
\end{equation}  
\begin{equation}
 \mathcal{L}_{\rm V} = \sum_{i}\{\bar{\psi}_i g_{iV}(\gamma^\mu V_{\mu})\psi_i - \frac{1}{2}m_V^2 V^\mu V_\mu \} \Theta(\bar{\psi}_i\psi_i) ,\label{e9}
\end{equation}
\begin{equation}
 \mathcal{L}_{\rm DIRAC} = b_4\frac{(g^2V_\mu V^\mu)^2}{4} ,\label{e12}
\end{equation}
where $\psi_i$ is the Dirac quark field, $B$ is the constant vacuum pressure, $m_V$ is the mass of the $V_\mu$ mesonic field, $g_{iV}$ is the coupling constant of the quark $i$ with the meson $V_\mu$, and $g$ = $g_{uV}$. The  $\Theta(\bar{\psi}_i\psi_i)$ is the Heaviside step function included to assure that the quarks exist only confined to the bag.

Applying MFA and the Euler-Lagrange equations, we can obtain the energy eigenvalue for the quark fields and the equation of motion for the mesonic $V_0$ field:
\begin{equation}
E_i =  \sqrt{m_i^2 + k_i^2} +g_{iV}V_0, \,
\label{eigen}   
\end{equation}
\begin{equation}
gV_0  + \bigg ( \frac{g}{m_v} \bigg)^2 \bigg (  b_4 (gV_0)^3 \bigg ) \nonumber 
= \bigg (\frac{g}{m_v} \bigg ) \sum_{i} \bigg (\frac{g_{iV}}{m_v} \bigg )n_i . \nonumber \\ \label{V0}
\end{equation}
To construct an EOS in MFA, we now consider the Fermi-Dirac distribution of the quarks, the Hamiltonian of the vector field, and the bag pressure value, $\mathcal{H} = -\langle \mathcal{L} \rangle$. We obtain
\begin{equation}
\varepsilon_i = \frac{\gamma_i}{2\pi^2}\int_0^{k_f} E_i ~k^2 dk,\label{nl4}
\end{equation}
\begin{equation}
\varepsilon =  \sum_i \varepsilon_i + B - \frac{1}{2}m_V^2V_0^2  - b_4\frac{(g^2V_0^2)^2}{4}, \label{nl5}
\end{equation}
with $\gamma_i=6~ (3~ {\rm colors} \times 2~ {\rm  spins})$ is the degeneracy factor.

To construct a hybrid star, the quark matter must lie outside the so-called stability window~\cite{Bodmer_1971, Witten_1984}, i.e., the energy per baryon must be higher than 930 MeV. Otherwise, the entire hybrid star would be converted into a strange star in a finite amount of time~\cite{Olinto1987, Lugones2015}. To accomplish this task, we use $B^{1/4} = 158$ MeV. We also define $G_V~\equiv~(g/m_V)^2$ and $X_V~\equiv~(g_{sV}/g_{uV})$.  The $X_V$ is then taken as $X_V = 0.4$ once its value was calculated based on symmetry group arguments ( { although it is worth to point out that the SU(3) symmetry here is also broken at mass level, as in the QHD case,} see reference \cite{Lopes_2021} for additional details). For the $G_V$, we use two different values to produce a weak and a strong interaction between the quarks, $G_V = 0.40$, and $0.42$ fm$^2$. The Dirac sea contribution is taken into account, assuming a small value for $b_4 \ (=0.5$). Within such a small value, we are still able to produce massive stars, with $M\geq2.0 M_\odot$, while assuring that the quark matter is energetically favorable at the limit of higher density. Finally, the pressure is easily obtained by thermodynamic relations: $p =
\sum_i \mu_i n_i - \varepsilon$.
\subsection{Maxwell construction and phase transition}
To identify the point where the hadronic matter becomes deconfined quark matter, we use the so-called Maxwell construction. In this approach, the transition pressure is the one where the Gibbs free energy per baryon $G/n_B$ of both phases intersect, being the
energetically preferred phase, the one with a lower
$G/n_B$~\cite{Chamel2013}. The Gibbs free energy per baryon coincides with the baryon chemical potential. Therefore, we call the intersection point critical pressure and critical chemical potential. The Maxwell criteria read:
\begin{equation}
\mu_H = \mu_Q = \mu_0, \quad \mbox{at} \quad p_H = p_Q = p_0 . 
\label{EL28}
\end{equation}
\begin{table}[t]
\centering
\begin{tabular}{cc|cccc}
\hline 
Hadron & Quark & $\mu_0$ & $p_0$& $\epsilon_H$ & $\epsilon_Q$ \\ \hline
Nucleonic & Weak& 1214 & 85 & 470 & 507\\
Nucleonic & Strong&1302& 129 & 557 & 609  \\
Hyperonic &  Weak  & 1228& 92 & 506 & 528  \\
Hyperonic& Strong & 1489 & 253 & 860 & 904  \\ \hline
\end{tabular} 
\caption{The critical chemical potential (in MeV), critical pressure (in MeV/fm$^3$), and the energy densities (in MeV/fm$^3$) for hadron and quark phases, within  Maxwell construction.} 
\label{T2}
\end{table}

The critical pressure ($p_0$), therefore,  marks the onset of the quark phase. We have a pure hadronic star for a neutron star whose central pressure is lower than $p_0$, and a hybrid star if $p_c~>~p_0$. When $p_c = p_0$ we have the minimum neutron star mass that presents quarks in the core.

We display in Tab.~\ref{T2} the critical chemical potential and the critical pressure for beta-stable matter with only Nucleons (Nucleonic) and with nucleons and hyperons (Hyperonic), as well with the quark matter with $G_V = 0.40$ fm$^2$ (Weak) and $G_V$ = 0.42 fm$^2$ (Strong). We also display the correspondent energy density for the hadronic and quark phases at the transition point. We can notice that by increasing the value of $G_V$, we also increase the critical chemical potential, pushing the hadron-quark phase transition to higher densities and pressures. In the same sense, the presence of hyperons softens the hadron EOS, making the hadron-quark phase transition more difficult. Indeed, for even higher values of $G_V$, we are not able to produce stable hybrid stars with hyperons once their critical chemical potentials are beyond those reached in the maximum mass.  These results are in accordance with the discussion presented in ref.~\cite{Lopes_ApJ}. 
\section{Spherically symmetric hybrid stars}
\subsection{Hydrostatic equilibrium}
%
In this section, we briefly review the theory of static equilibrium distribution of matter, which is spherically symmetric but whose stress tensor is, in general, locally anisotropic, as originally introduced in ref.~\cite{Bowers_1974} and called here as BL model.  In the isotropic case, the stress-energy tensor reads: $T_{\mu\nu}$ = diag($\rho, -p, -p, -p$). We now introduce anisotropies without breaking the spherical symmetry by assuming the following stress-energy tensor: $T_{\mu\nu}$ = diag($\rho, -p_r, -p_t, -p_t$). Spherical symmetry implies that (in canonical coordinates) the stress-energy tensor $T_{\mu\nu}$ is diagonal, and moreover, $p_\phi =p_\theta = p_t$~\cite{DelgadoeDelgado2018}.

We can now redefine the anisotropic energy-momentum tensor as done in refs. ~\cite{Doneva_2012, Silva_2015, DelgadoeDelgado2018}:
\begin{align}
    T_{\mu\nu} = (\rho+p_t)u_\mu u_\nu + (p_r-p_t) k_\mu k_\nu + p_t g_{\mu\nu},
    \label{eq:tmunu_aniso}
\end{align}
where $\rho$, $p_r$, and $p_t$ are the energy density, radial pressure, and tangential pressure, respectively. $k_\mu$ is the unit radial vector ($k^\mu k_\mu = 1$) with $u^\mu k_\mu = 0$. The Schwarzschild metric for this type of star having the spherically symmetric and static configuration is defined as 
\begin{align}
    ds^2= - e^{2 \nu}dt^2+e^{2 \lambda}dr^2+r^2d\theta^2-r^2 \sin^2\theta  d\varphi^2\,, \label{Scz}
\end{align} 
where $r$, $\theta$, and $\phi$ are the Schwarzschild coordinates. 

Applying Einstein's field equations, we obtain:
\begin{align}
\lambda'(r) = \frac{8\pi r^2 e^{2\lambda(r)}\rho(r) - e^{2\lambda(r)} +1}{2r}, \label{elambda}
\end{align}
\begin{align}
\nu'(r) = \frac{8\pi r^2 p_r(r) e^{2 \lambda(r)}+e^{2\lambda(r)}-1}{2r}, \label{enu}
\end{align}
and the contracted Bianchi identities give us the following:
\begin{align}
\frac{dp_r}{dr} = - (\rho + p_r)\frac{\nu'}{2} + \frac{2}{r}(p_t - p_r). \label{edp}
\end{align}

Finally, by isolating the $\nu'$ in Eq.~\ref{enu} and replacing it in Eq.~\ref{edp}, we can write the equilibrium equations in the Tolman-Oppenheimer-Volkoff form~\cite{Doneva_2012}:
\begin{align}
    \frac{dp_r}{dr}=-\frac{\left( \rho + p_r \right)\left(m + 4\pi r^3 p_r \right)}{r\left(r -2m\right)} +\frac{2\sigma}{r} \,,
    \label{tov1:eps}
\end{align}
\begin{align}
    \frac{dm}{dr}=4\pi r^{2}\rho\,,
    \label{tov2:eps}
\end{align}
where $\sigma=p_t-p_r$ is the anisotropy parameter, `$m$' is the mass enclosed within the radius $r$. The radial pressure is then obtained from a pre-determined EOS. On the other hand, for the case of the transverse pressure, we use the BL model in the following \cite{Bowers_1974}: 
\begin{align}
    \label{Anisotropy_eos}
    p_t = p_r + \frac{\lambda_{\rm BL}}{3} \frac{(\rho+3p_r)(\rho + p_r)r^2}{1-2m/r} \,,
\end{align}
where the factor $\lambda_{\rm BL}$ measures the degree of anisotropy in the fluid. There are some boundary conditions required to solve the above Eqs. (\ref{tov1:eps}-\ref{Anisotropy_eos}) as done in Refs. \cite{Biswas_2019, Das_ILC_2022}. Also, different fluid conditions must be satisfied for the anisotropic stars, such as (i) $p_r, p_t$, and $\rho > 0$, (ii) $0<c_{\rm s, t}^2<1$ (where $c_{\rm s, t} =dp_t/d\rho$), (iii)  $p_r = p_t$ for $r =0$, etc. Other conditions are mentioned in ref.~\cite{Das_ILC_2022}. { By construction, the surface of the star is defined at the point that the radial pressure vanishes. For hadronic and hybrid stars, this coincides with the energy density going to zero, which implies a zero tangential pressure as well (see Eq.~\ref{Anisotropy_eos}). Additional discussion about it can be seen in the appendix.} { The interest in the BL model has increased in the last few years, and it has been applied to different goals. In ref.~\cite{Das_ILC_2022,Biswas_2019,Das2024JCAP} it was applied to study I-Love-C relations and tidal deformability; in ref.~\cite{Rahmansyah_2020} it was used to study the presence of hyperons in anisotropic stars; in ref.~\cite{Lopes2023M,Lopes2024EPJC} it was used to study strange stars; in ref.~\cite{Pretel2024EPJC} it was employed to study equilibrium and collapse of anisotropic neutron stars; in ref.~\cite{Pattersons_2021} the BL model was expanded to study slowly rotating anisotropic neutron stars; finally in ref.~\cite{Jyo2024EPJC} the model was used to study the effects of anisotropies in dark energy stars.}

\begin{figure*}[t]
\begin{tabular}{ccc}
\centering 
\includegraphics[scale=.48, angle=0]{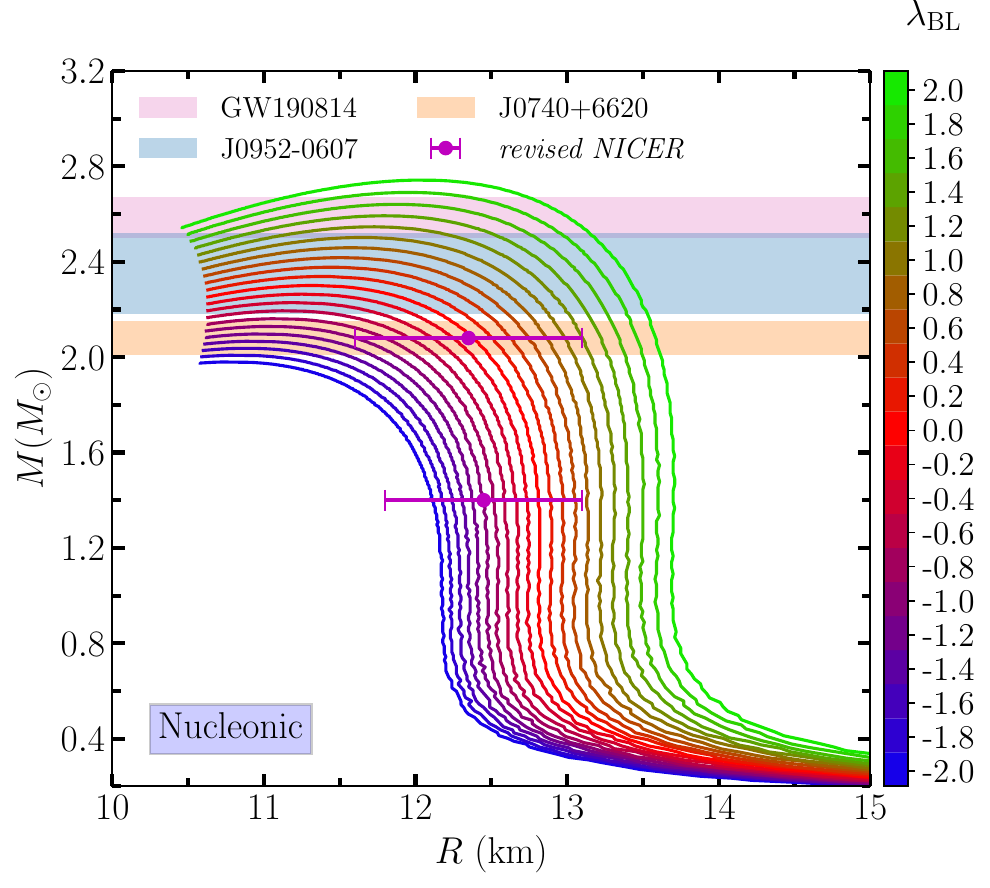} &
\includegraphics[scale=.48, angle=0]{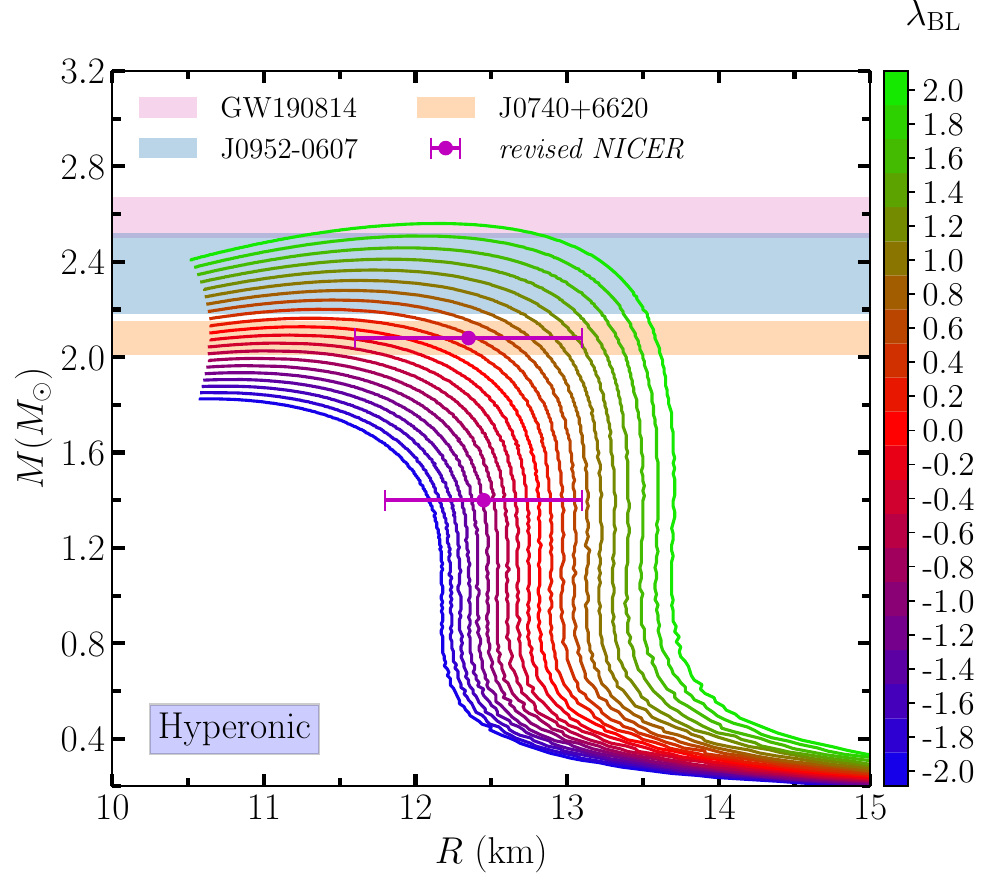} \\
\includegraphics[scale=.48, angle=0]{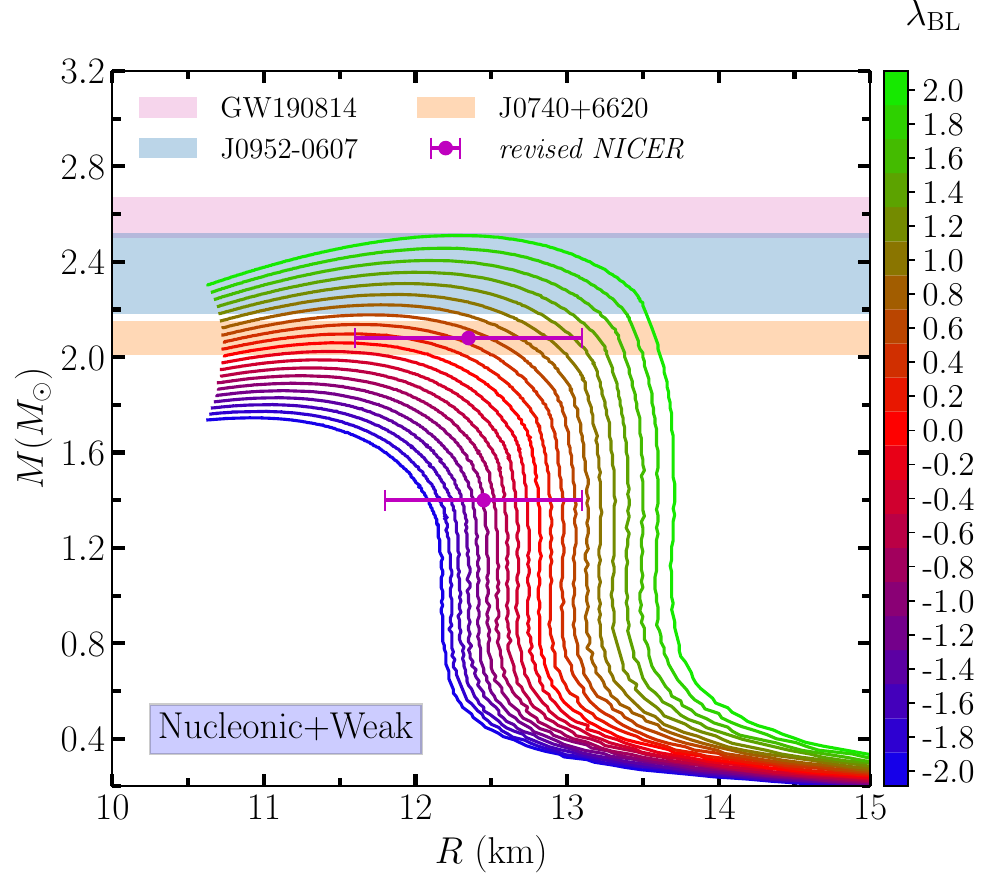} &
\includegraphics[scale=.48, angle=0]{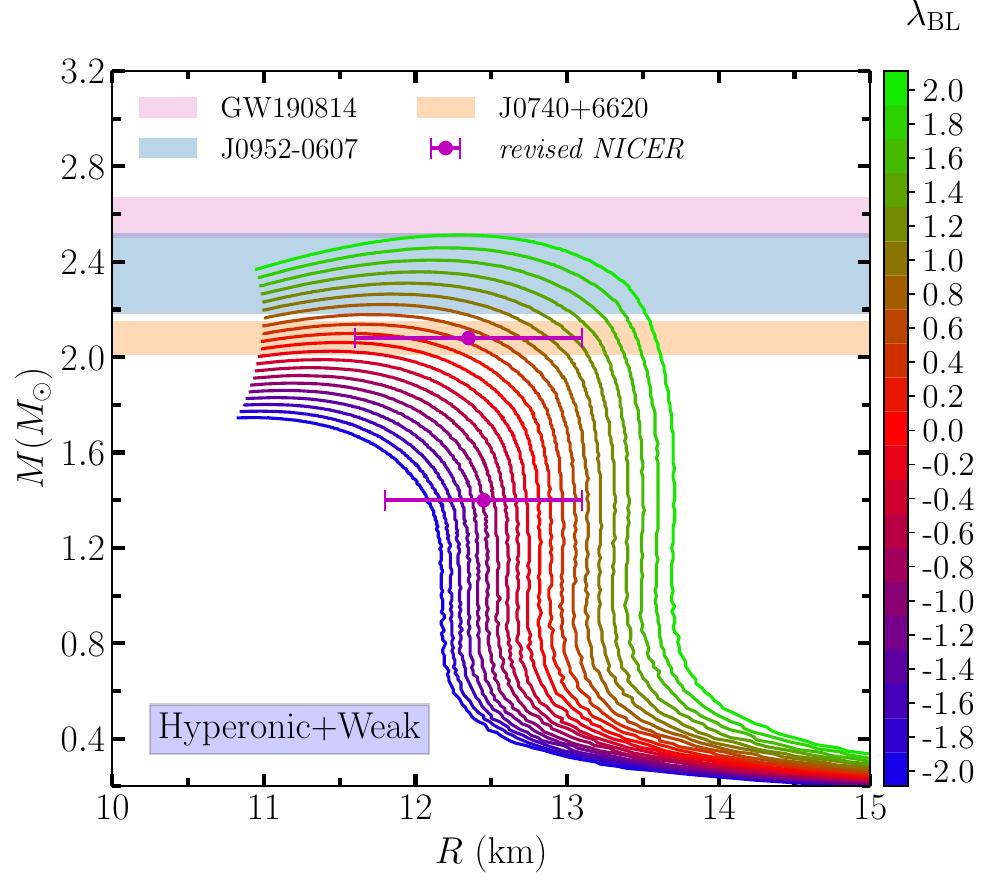}\\
\includegraphics[scale=.48, angle=0]{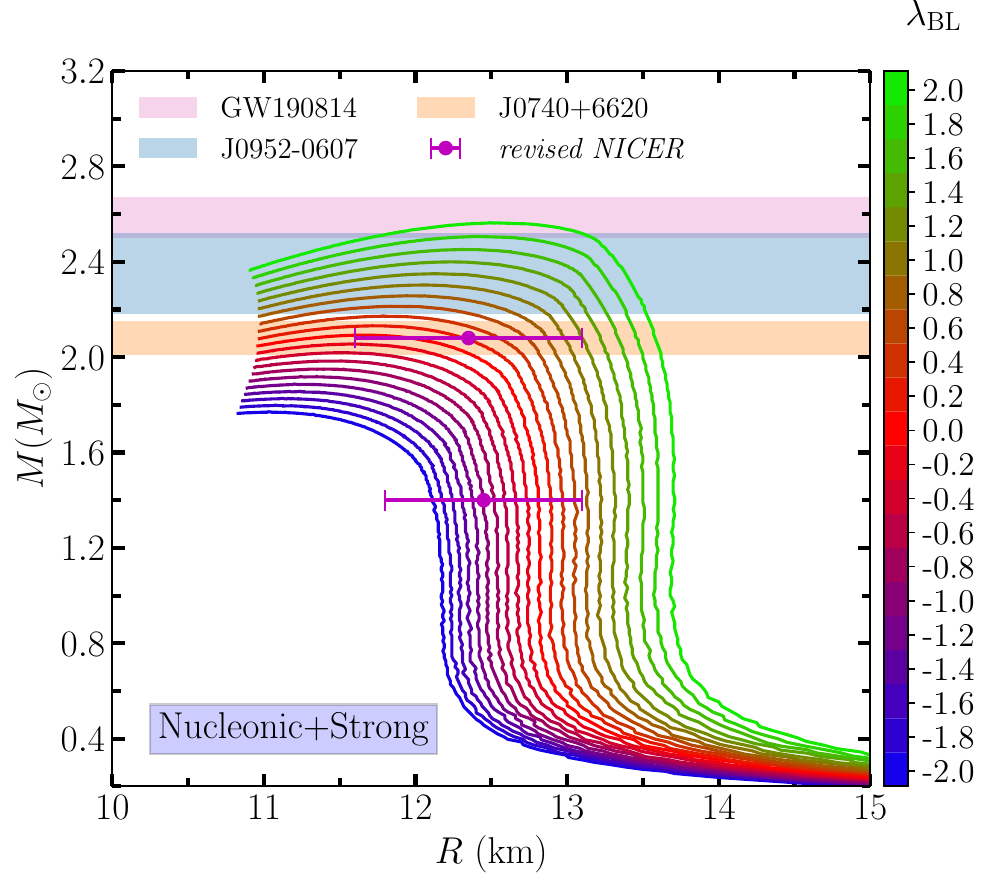} &
\includegraphics[scale=.48, angle=0]{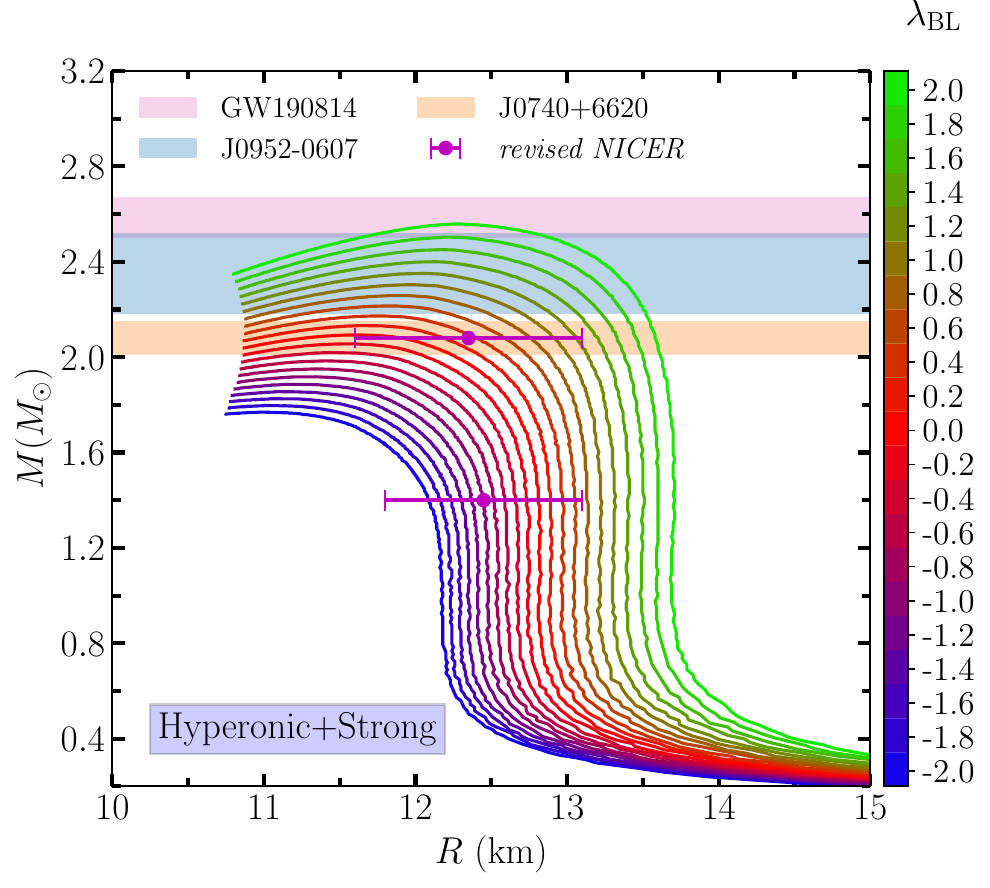}\\
\end{tabular}
\caption{Mass-radius relations for different neutron stars' configurations. The anisotropic parameter lies $-2.0~<~\lambda_{\rm BL}~<~+2.0$. } \label{FL1}
\end{figure*}

In Fig.~\ref{FL1}, we display the mass-radius relation for different values of $\lambda_{\rm BL}$ with some neuron star configurations: the neutron stars that present hyperons are on the right, while those that do not present hyperons are on the left. From top to bottom, there are pure hadronic stars, hybrid stars with the weak quark-quark interaction, and hybrid stars with the strong quark-quark interaction. { For all models, we use the BPS EOS~\citep{BPS} for the neutron star's outer crust and the BBP EOS~\citep{BBP} for the inner crust.  We use the BPS+BBP EoS up to the density of 0.0089 fm$^{-3}$  and from this point on, we use the QHD EoS, as suggested in ref.~\cite{Glenbook}.  Imposing  $p_{\rm crust} = p_{\rm core}$ at the crust-core transition, we found that the core EoS begins around 0.03 fm$^{-3}$. This procedure is the same as done in refs.~\cite{Lopes2023PRDb,Lopes2023PTEP}. }

\begin{figure*}[t]
\begin{tabular}{ccc}
\centering 
\includegraphics[scale=.48, angle=0]{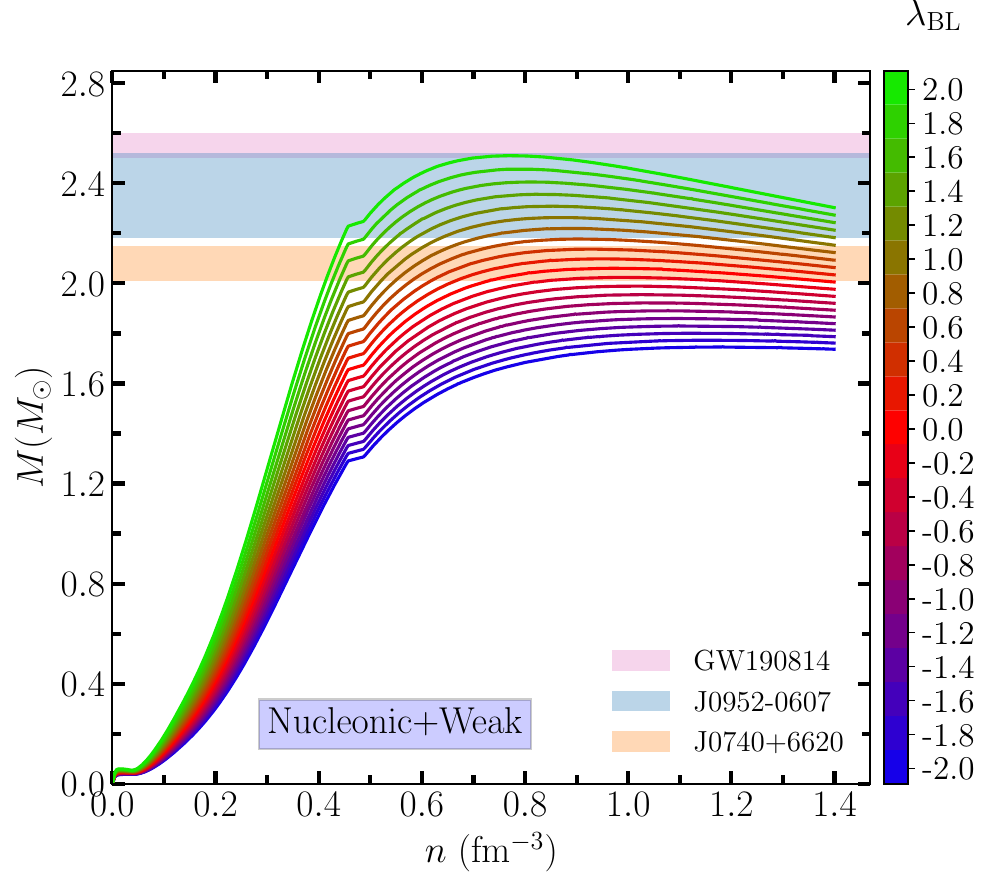} &
\includegraphics[scale=.48, angle=0]{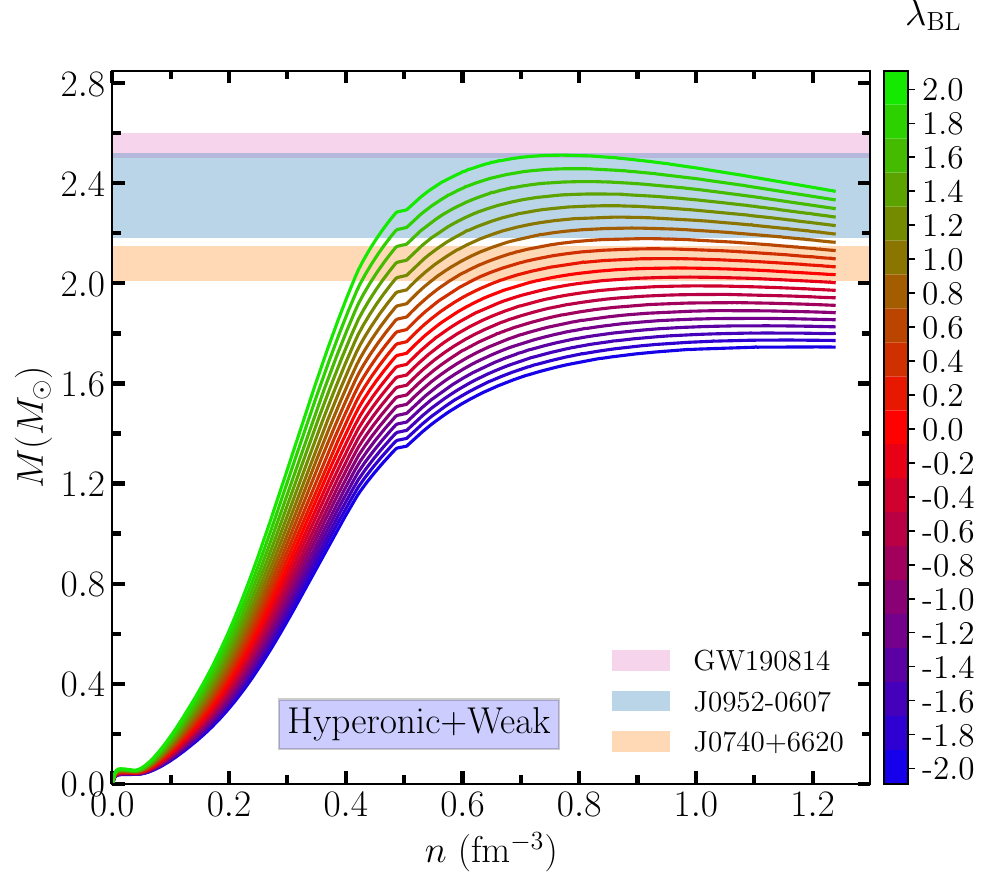}\\
\includegraphics[scale=.48, angle=0]{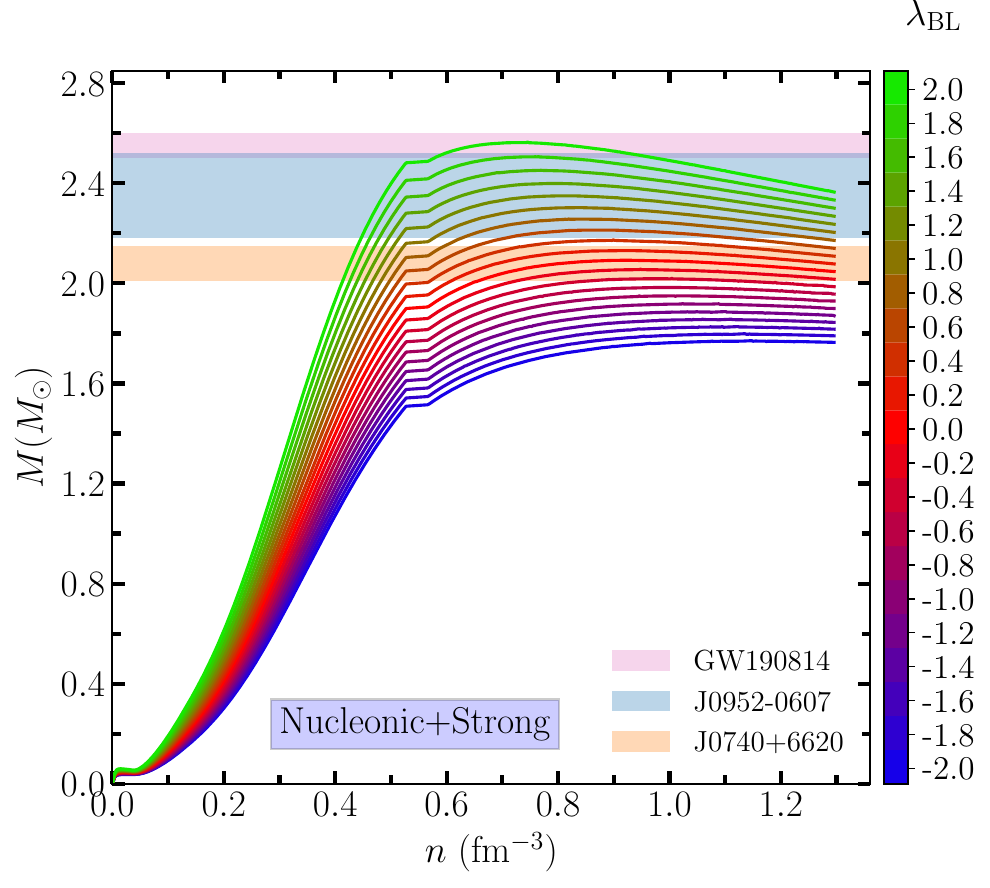} &
\includegraphics[scale=.48, angle=0]{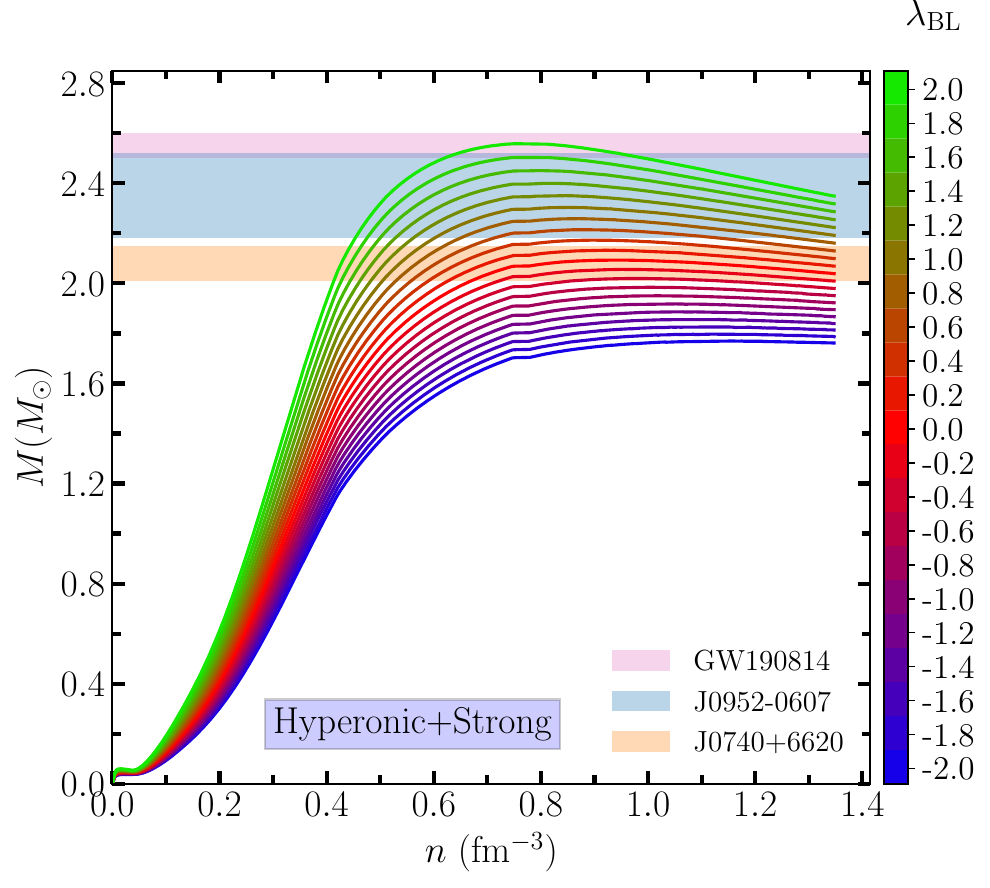}\\
\end{tabular}
\caption{Hybrid star masses as a function of the central density. The anisotropic parameter lies $-2.0~<~\lambda_{\rm BL}~<~+2.0$. } \label{FL2}
\end{figure*}

We follow~\cite{Lopes2023M} and begin our analyses with the symmetric case. For $\lambda_{\rm BL}$ = 0, we see that the maximum mass is 2.30$M_\odot$ for a pure Nucleonic star, while its central number density is around 0.94 fm$^{-3}$. A maximum mass of 2.13$M_\odot$, with a central density around 0.99 fm$^{-3}$, was found for a hadronic neutron star with nucleons and hyperons. In the case of hybrid neutron stars, the maximum mass is 2.06 $M_\odot$ for the weak quark-quark interaction and 2.09 $M_\odot$ for the strong quark-quark interaction. It is interesting to point out that we obtain the same maximum mass regardless of whether the hadronic EOS presents hyperons. Such a feature was first pointed out in ref.~\cite{Lopes_ApJ}. Concerning their central densities, we see that for the weak quark-quark interaction, we have a central density around 0.95 fm$^{-3}$ for a pure Nucleonic hybrid star and around 0.92 fm$^{-3}$ for a hybrid star with nucleons and hyperons. For a strong quark-quark interaction, we have exactly the opposite: a central density around 0.92 fm${-3}$ for a pure Nucleonic hybrid star and around 0.95 fm$^{-3}$ for a hybrid star with nucleons and hyperons.   Nevertheless, although the presence of hyperons does not affect the maximum mass of hybrid stars, their presence makes the phase transition more difficult, pushing it to higher densities. This can more easily seen from Fig.~\ref{FL2}, where we display the hybrid star mass as a function of the central density $n$. The onset of the quark matter can be identified by the gap in the number density (represented by small horizontal lines, indicating that the mass does not change). The first hybrid star for a weak quark-quark interaction has a mass of 1.63 $M_\odot$, and a density around 0.48 fm$^{-3}$ for a Nucleonic EOS and $M = 1.68~M_\odot$ with $n = 0.50$ fm$^{-3}$ for a Hyperonic hybrid star. In the case of the strong quark-quark interaction, we have the first hybrid star with a mass of 1.84 $M_\odot$ with $n = 0.57$ fm$^{-3}$ for a Nucleonic hybrid star and $M =$ 2.05 $M_\odot$ with $n = 0.78$ fm$^{-3}$ for a Hyperonic one. Now we discuss the radius of the 1.4 $M_\odot$ canonical star. For the isotropic case, we found that all stars have the same radius of 12.82 km, as neither hyperons nor quarks are present at such low mass.

We now discuss the role of anisotropy on the neutron star properties. For a pure Nucleonic star, the maximum mass drops to 1.98 $M_\odot$ for $\lambda_{\rm BL}$ = -2.0 while  peaks 2.74 for$\lambda_{\rm BL}$ = +2.0. On the other hand, the central density is as high as 1.12 fm$^{-3}$ or $\lambda_{\rm BL}$ = -2.0 and drops to 0.77 fm$^{-3}$ for $\lambda_{\rm BL}$ = +2.0. For a pure Hyperonic neutron star we have $M = $ 1.83 $M_\odot$ for $\lambda_{\rm BL}$ = -2.0, and $M = $ 2.56 for $\lambda_{\rm BL}$ = +2.0. In the same sense, we have $n = 1.20$ fm$^{-3}$ for $\lambda_{\rm BL}$ = -2.00, and $n =0.79$ fm$^{-3}$ for$\lambda_{\rm BL}$ = +2.0. As can be seen, the presence of a positive anisotropy increases the maximum mass but reduces the central density and vice-versa.

In the case of hybrid stars with a weak quark-quark interaction, we found a maximum mass of 1.75 $M_\odot$ for $\lambda_{\rm BL}$ = -2.0 and $M =$ 2.51 for  $\lambda_{\rm BL}$ = +2.0. Here, as in the isotropic case, the maximum mass does not depend on the presence of hyperons or not. Indeed, we have the same maximum mass for a fixed $\lambda_{\rm BL}$ and a fixed quark EOS. The central density presents a small difference, $n = 1.18$  fm$^{-3}$ for Nucleonic+Weak hybrid star, and $n = 1.19$  fm$^{-3}$ for a Hyperonic+Weak hybrid one for $\lambda_{\rm BL}$ = -2.0. For $\lambda_{\rm BL}$ = +2.0 we have $n = 0.77$ fm$^{-3}$ and $n = 0.76$ fm$^{-3}$ for a Nucleonic and a Hyperonic hybrid star respectively. { This can be, nevertheless, a model-dependent feature.}

For a strong quark-quark interaction, we again have the maximum mass independent of the presence of hyperons for all values of $\lambda_{\rm BL}$. The maximum mass lies between 1.77 $M_\odot$ for $\lambda_{\rm BL}$ = -2.0 up to 2.56 $M_\odot$ for $\lambda_{\rm BL}$ = +2.0. It is interesting to notice that the maximum mass for a strong quark-quark interaction hybrid star presents the same maximum mass as the pure hadronic Hyperonic star for $\lambda_{\rm BL}$ = +2.0. For the other value of $\lambda_{\rm BL}$, the maximum mass of the hybrid star is always below the one of the hadronic Hyperonic star. In the case of the central density, we find values between 1.15 fm$^{-3}$ for $\lambda_{\rm BL}$ = -2.0 and 0.74 fm$^{-3}$ for $\lambda_{\rm BL}$ = +2.0 for the case Nucleonic+Strong. For Hyperonic+Strong, we have values between 1.16 fm$^{-3}$ and 0.74 fm$^{-3}$ for  $\lambda_{\rm BL}$ = -2.0 and $\lambda_{\rm BL}$ = +2.0 respectively. It is important to point out that for $\lambda_{\rm BL}$ = +2.0, the central density is below the density at the quark-hadron phase transition. Therefore, we do not have a stable hybrid star for this configuration. All stable stars are pure hadronic. This is why both Hyperonic and Hyperonic+Strong have the same mass for $\lambda_{\rm BL}$ = +2.0.

We can also discuss what is the minimum neutron star mass that presents quarks in their core. We see that for $\lambda_{\rm BL}$ = -2.0 we have 1.23 $M_\odot$ (Nucleonic+Weak), 1.27 $M_\odot$ (Hyperonic+Weak),  1.49 $M_\odot$ (Nucleonic+Strong) and  1.67 $M_\odot$ (Hyperonic+Strong). For $\lambda_{\rm BL}$ = +2.0, we have 2.21 $M_\odot$ (Nucleonic+Weak), 2.28 $M_\odot$ (Hyperonic+Weak), and 2.47 $M_\odot$ (Nucleonic+Strong). Hyperonic+Strong does not have a stable hybrid star. 

\begin{figure}[t]
  \begin{centering}
\includegraphics[width=0.45\textwidth,angle=0]{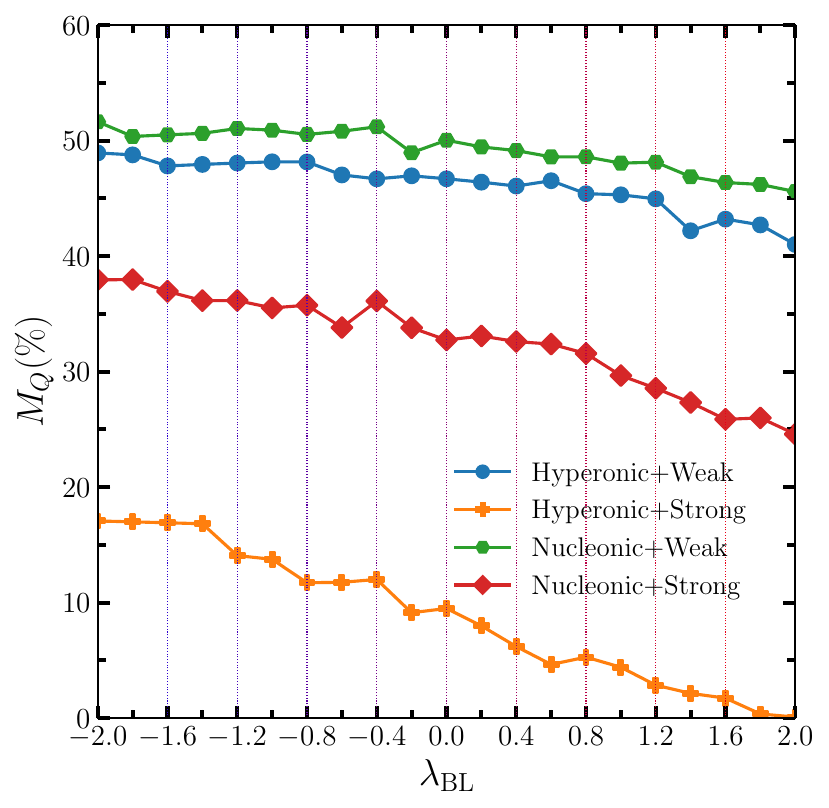} \\
\caption{Percentual amount of quarks in the core of the most massive hybrid stars as a function of  $\lambda_{\rm BL}$ for different hybrid star configurations as discussed in the text. } \label{Fmq}
\end{centering}
\end{figure}

We now estimate the percentual amount of quark present in the maximally massive hybrid star for each neutron star configuration. To accomplish that, we follow refs. ~\cite{LOPES2021NPA, Lopes_2022} and solve the hydrostatic equilibrium equations for the quark EoS from the density corresponding to the central density at the maximum mass neutron stars and stop at the density corresponding to the critical chemical potential. The results are presented in Fig.~\ref{Fmq}. We see that the configuration that presents a higher amount of quarks in the core is Nucleonic+Weak, since it favors the earlier onset of quark matter, but still yields a large maximum mass. In the same sense, the lower values of quark mass are within the Hyperonic+Strong configuration. We also see that, despite some fluctuations, the percentage of quarks decreases with the increase of $\lambda_{\rm BL}$. For $\lambda_{\rm BL}$ = -2.0, the total amount of quarks are 52$\%$ for Nucleonic+Weak, 49$\%$ for Hyperonic+Weak, 38$\%$ for Nucleonic+Strong, and 17$\%$ for Hyperonic+Strong. For $\lambda_{\rm BL}$ = +2.0, these values drop to 46$\%$ for Nucleonic+Weak, 41$\%$ for Hyperonic+Weak, 25$\%$ for Nucleonic+Strong, and 0$\%$ for Hyperonic+Strong, once the maximum mass is reached before the hadron-quark phase transition. The decrease of the percentual amount of quarks with $\lambda_{\rm BL}$ can be explained by the reduction of the central density with $\lambda_{\rm BL}$ as well.

We now discuss the radius of the canonical 1.4 $M_\odot$ and some constraints related to pulsars' observations. For all hybrid stars, the radius of the canonical star is almost independent of internal composition. It varies from 12.09 km for $\lambda_{\rm BL}$ = -2.0, up to 13.71 km for $\lambda_{\rm BL}$ = +2.0, independently from both: the presence of hyperons, as well the strength of the quark-quark interaction.  Concerning the constraint related to the canonical star, we see that we need a value between -2.0 $\leq~\lambda_{\rm BL}~\leq$+0.6 to satisfy the refined NICER constraint, $11.80 <R_{1.4}< 13.10$ km ~\cite{Miller_2021}. About the PSR J0740+6620 with a mass of 2.08 $\pm$ 0.07 $M_\odot$ and a radius of 12.35 $\pm$ 0.75 km ~\cite{Miller_2021}, we see that all hybrid stars configurations satisfy this constraint if we assume that  -0.2 $\leq~\lambda_{\rm BL}~\leq$+1.2. Such restriction is independent of the presence of hyperons or the strength of the quark-quark interaction. Combining both constraints, we see that -0.2 $\leq~\lambda_{\rm BL}~\leq$+0.6 produce hybrid stars that simultaneously satisfy the canonical 1.4 $M_\odot$ and the revised mass and radius value of the PSR J0740+6620.

About the less orthodox constraints, we see that the speculative mass of the black widow pulsars PSR J0952-0607, $M = 2.35\pm0.17 M_\odot$ \cite{Romani_2022}, can be reached for $\lambda_{\rm BL}~\geq$ +0.4 for a strong quark-quark interaction and $\lambda_{\rm BL}~\geq$ +0.6 in the case of the weak quark-quark interaction. Here again, the results are independent of the presence of hyperons.  The mass-gap object in the GW190814 event~\cite{RAbbott_2020}, whose mass was estimated to be $2.50 - 2.67 \ M_\odot$ can be reached for $\lambda_{\rm BL}~\geq~+1.8$ for the strong case, and only within $\lambda_{\rm BL} = +2.0$ in the weak one. Finally, the constraint related to the HESS object~\cite{Doroshenko_2022} cannot be fulfilled for any value of $\lambda_{\rm BL}$.
\subsection{Tidal deformation}
\begin{figure*}[tb]
\begin{tabular}{ccc}
\centering 
\includegraphics[scale=.48, angle=0]{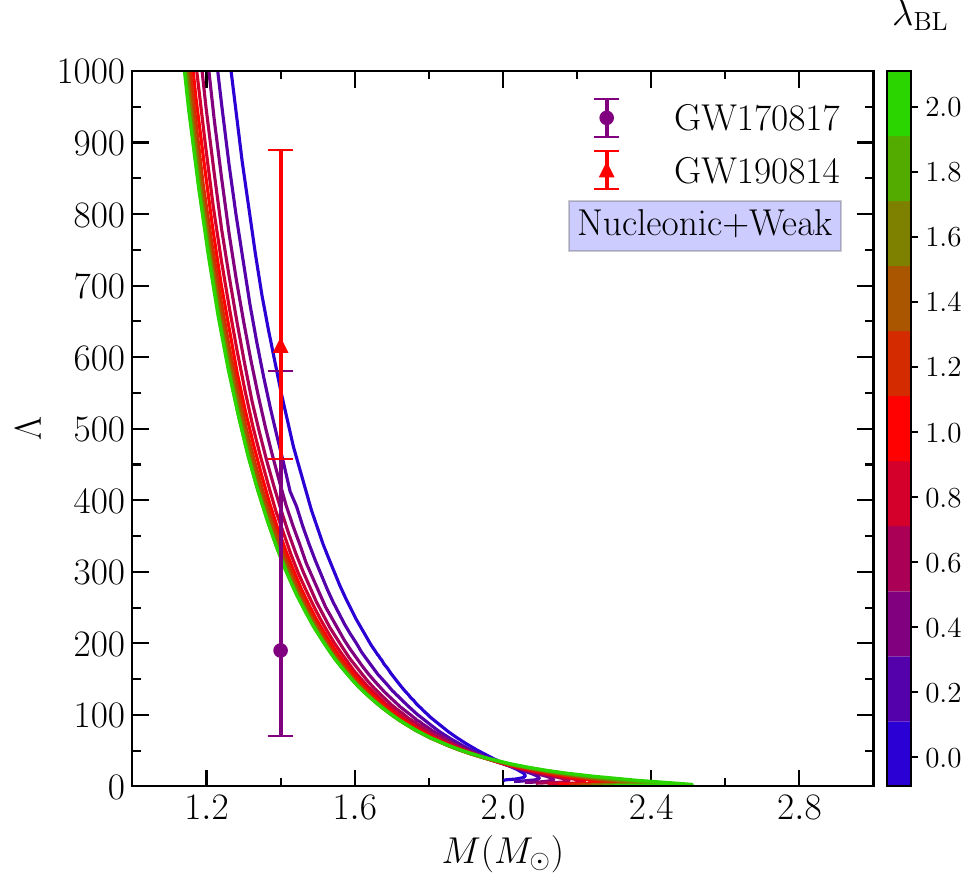} &
\includegraphics[scale=.48, angle=0]{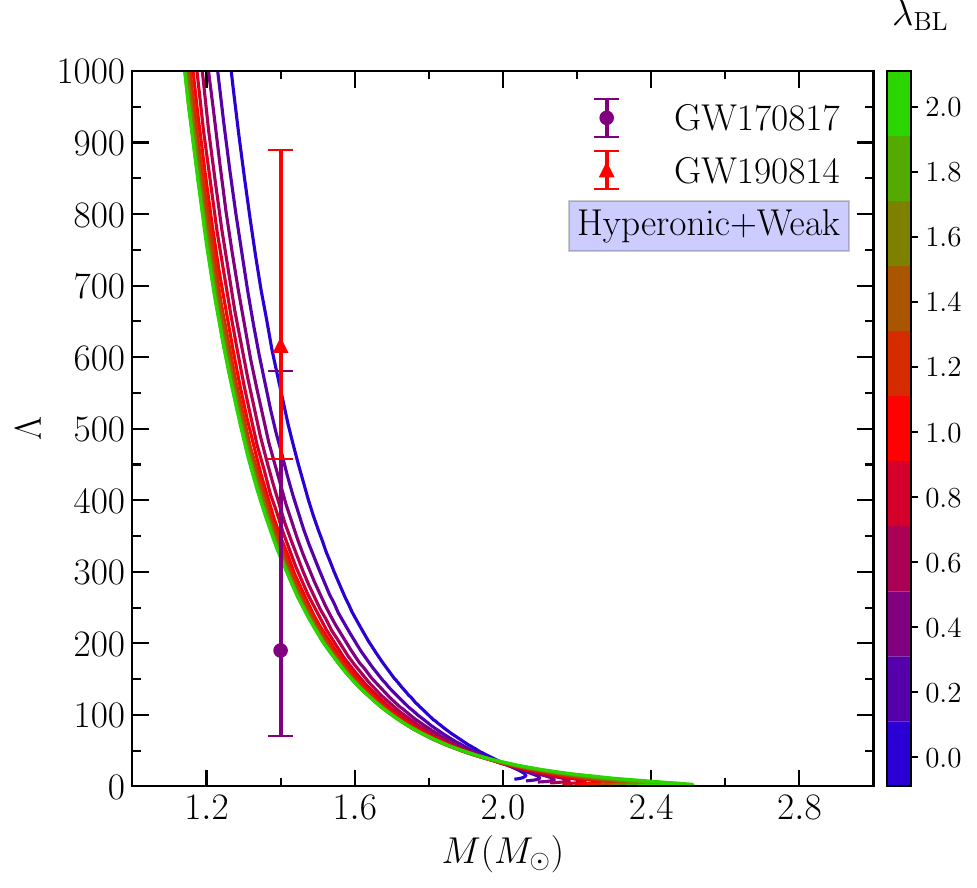}\\
\includegraphics[scale=.48, angle=0]{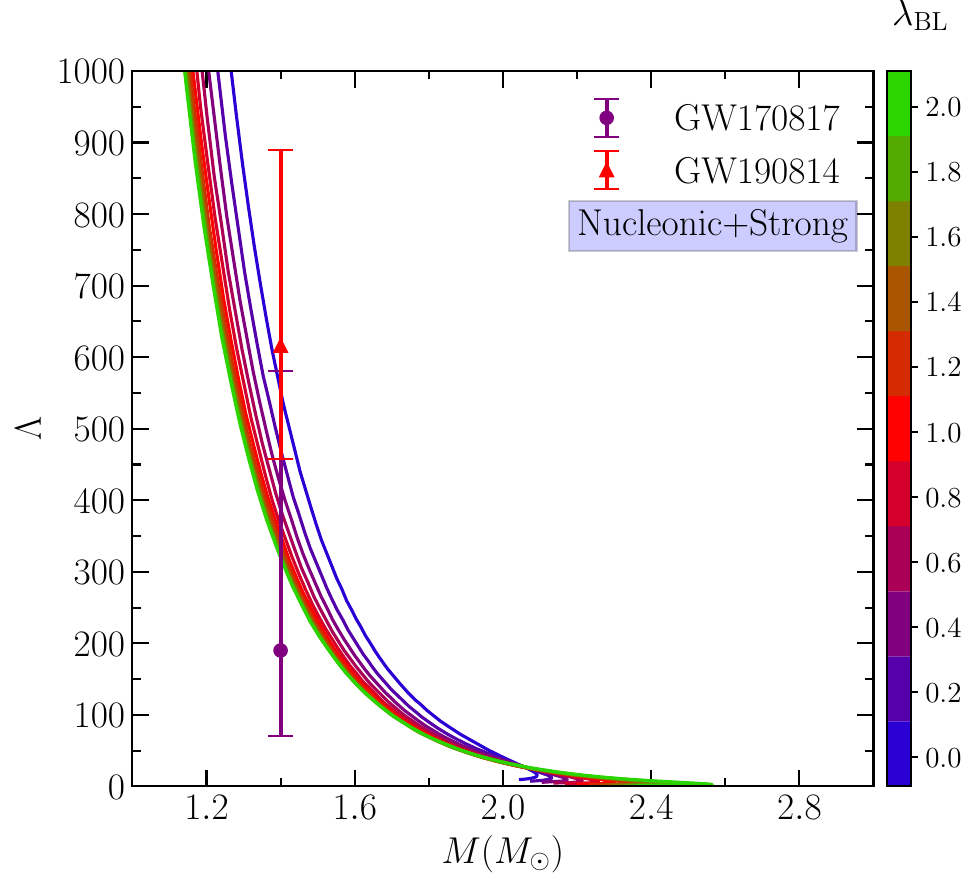} &
\includegraphics[scale=.48, angle=0]{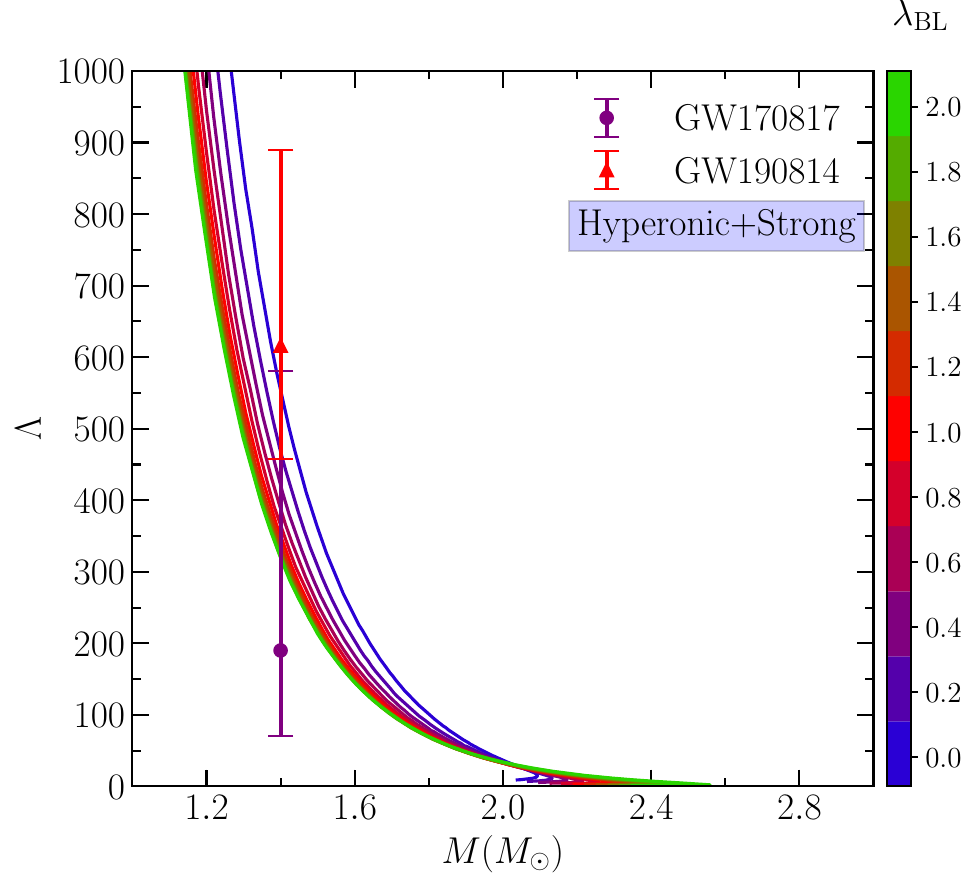}\\
\end{tabular}
\caption{Dimensionless tidal parameter for different configurations of hybrid stars. The anisotropic parameter lies $0.0~<~\lambda_{\rm BL}~<~+2.0$.}
\label{FL4}
\end{figure*}
Tidal deformability ($\Lambda$) is a quantity that measures the amount of deformation in the shape of the star when it is in the field of its companion star. Already, LIGO/VIRGO detected the binary neutron star deformation in 2017. It has a unique relationship with the tidal Love number ($k_2$) and the compactness $(C)$ of the star~\cite{Hinderer_2008, Hinderer_2009}:
\begin{equation}
    \Lambda = \frac{2}{3} k_2 C^{-5}
\end{equation}

One can estimate the magnitude of $k_2$ by solving the following differential equation in the Throne, and Campolattaro metric \cite{Throne_1967} for an anisotropic star with an EOS is given by ~\cite{Biswas_2019, Ray2023Nov}
\begin{align}
    H^{''}(r) + A H^{'}(r) + B H(r) = 0 \ .
\end{align}
{ With $y = rH^{'}/H$, we can rewrite the equation as follow as 
\begin{align}
    r y^{'}(r) + y(r)^2 + (rA-1) y(r) + r^2 B = 0 \ \,
\end{align}
}
where, $A$ and $B$ are represented as 
\begin{align}
    &A = - \left(\frac{- e^{2\lambda (r)} -1}{r} - 4\pi r  e^{2 \lambda(r)} \left(p_r(r) - \rho_r(r)\right) \right), \\
    &B = - \frac{dp_r}{dp_t}\Bigg(\frac{4 e^{2\lambda(r)}+e^{4\lambda(r)}+1}{r^2} + 64 \pi^2 r^2 p_r^2(r) e^{4\lambda (r)} \nonumber \\
    &+ 16 \pi e^{2\lambda(r)}\left\{p_r(r) (e^{2\lambda(r)} -2) - p_t(r) - \rho(r)\right\} - 4 \pi c_{s,t}^2 \nonumber \\
    & \times e^{2\lambda(r)} (p_r(r) + \rho(r)) - 4\pi e^{2\lambda(r)}(p_r(r) +\rho(r)) \Bigg) \, ,
\end{align}
where { $c_{s,t}^2 = dp_t/d\rho$ is the squared speed of sound}. The term $dp_r/dp_t$ can be obtained from Eq. \ref{Anisotropy_eos} for a particular star with a fixed value of $\lambda_{\rm BL}$. { The differential equation can be solved with the boundary condition $y(0)=2$ for a fixed central density \cite{Damour_2009, Hinderer_2008}. The solution at the surface of the star ($y_R$) can be taken to obtain the tidal Love number as mentioned in the following ~\cite{Hinderer_2008, Hinderer_2009, Flores_2020, DasBig_2021}.  
\begin{align}
    k_2 &= \, \frac{8}{5} C^5 (1-2C)^2 \big[ 2(y_R-1)C - y_R + 2 \big]
    \nonumber \\ 
    &\times \Big\{ 2C \big[ 4(y_R+1)C^4 + 2(3y_R-2)C^3 - 2(11y_R-13)C^2 
    \nonumber \\ &
    + 3(5y_R-8)C - 3(y_R-2) \big]+ 3(1-2C)^2 
    \nonumber \\ &
    \times \big[ 2(y_R-1)C-y_R+2 \big] \log(1-2C) \Big\}^{-1} \, .
    \label{eq:k2}
\end{align}
}

In Fig. \ref{FL4}, we depict the result for the tidal deformability as the function of mass for four cases only for positive $\lambda_{\rm BL}$ values. { We chose only the positive values because the speed of sound becomes negative for negative values of $\lambda_{\rm BL}$ for NSs as well as hybrid stars. The unphysical solution of the speed of sound gives obscure solutions for $\lambda_{\rm BL}$. The detailed discussion can be found for NSs in refs. \cite{Biswas_2019, Das_ILC_2022}.}

From the figure, we observe that the neutron stars with larger radii, i.e., larger values of $\lambda_{\rm BL}$, actually have lower values of $\Lambda$. This is the opposite when we compare pure isotropic within different models. For instance, in ref.~\cite{Lopes2023PRDb}, the authors vary the slope of the symmetry energy and found that when the radius of the canonical star is 12.58 km, we have $\Lambda_{1.4}$ = 515; but when $R_{1.4} = 14.30$ km, $\Lambda_{1.4} = 812$, an expected behavior. But when anisotropies are taken into account, for $R_{1.4} = 12.82$ km we have $\Lambda_{1.4} = 523$ for $\lambda_{\rm BL} = 0.0$,  $R_{1.4} = 13.22$ km with $\Lambda_{1.4} = 349$ for $\lambda_{\rm BL} = +1.0$ and finally, $R_{1.4} = 13.71$ km with $\Lambda_{1.4} = 320$ for $\lambda_{\rm BL} = +2.0$. This behavior was already pointed out in ref.~\cite{Das_ILC_2022}, and it is because the love number $k_2$ decreases with $\lambda_{\rm BL}$ (see fig. 5 from ref.~\cite{Das_ILC_2022}).
We can also see that the values for $\Lambda_{1.4}$, as the $R_{1.4}$, are the same for the four configurations as there are no hyperons or quarks at this mass value. The dimensionless tidal parameter for the maximally massive stars is almost independent of the internal composition. It is around 15-16 for $\lambda_{\rm BL} = 0.0$ and about 2-3 for $\lambda_{\rm BL} = +2.0$.

Concerning the gravitational wave constraints, we see that we easily satisfy the observation related to the GW170817,  $70<\Lambda_{1.4} <580$.~\cite{Abbott_2018}, once the tidal parameter decreases with $\lambda_{\rm BL}$. About the mass-gap object in the GW190814 event, the constraint $458 <\Lambda_{1.4}<889$~\cite{RAbbott_2020} can also be fulfilled if $\lambda_{\rm BL}~\leq$ 0.2.
\subsection{Moment of Inertia}
\begin{figure*}[t]
\begin{tabular}{ccc}
\centering 
\includegraphics[scale=.48, angle=0]{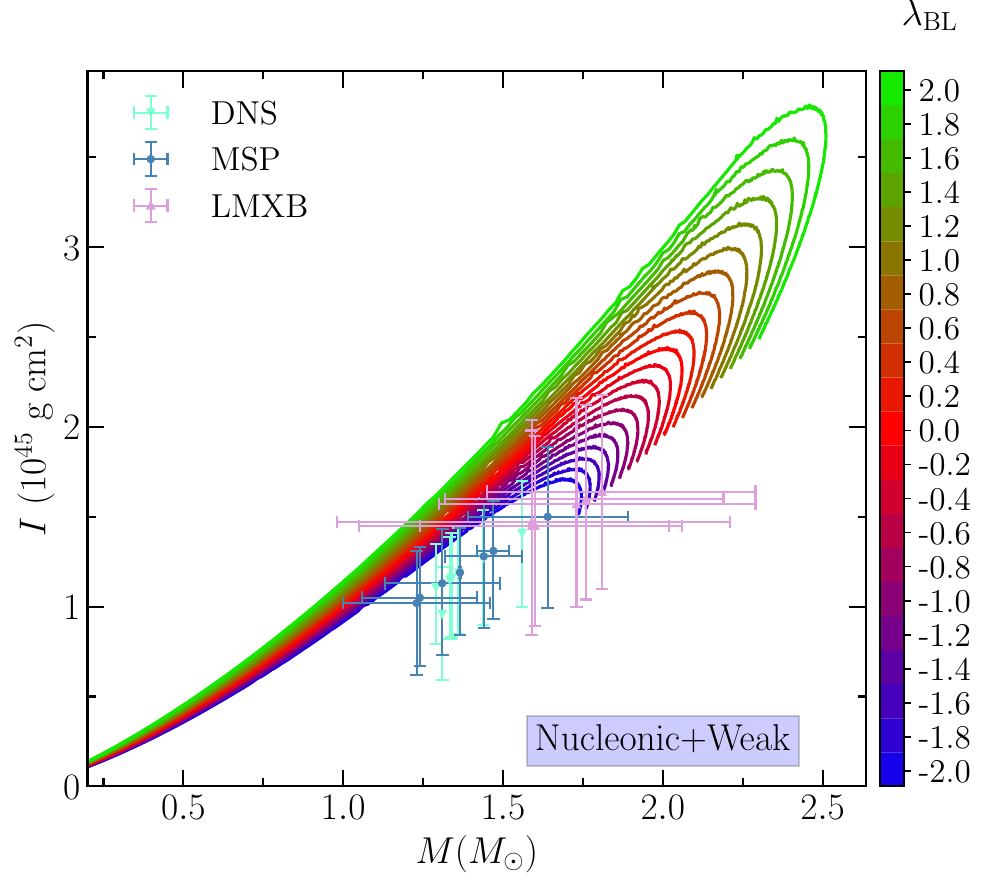} &
\includegraphics[scale=.48, angle=0]{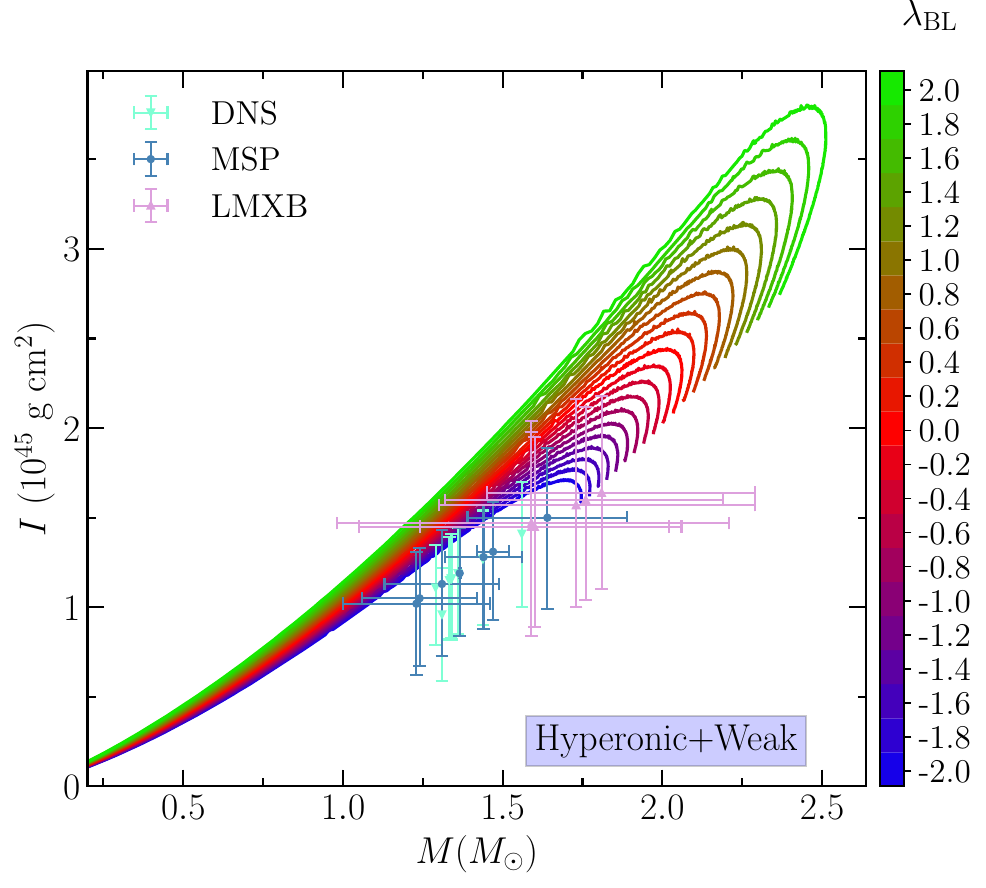}\\
\includegraphics[scale=.48, angle=0]{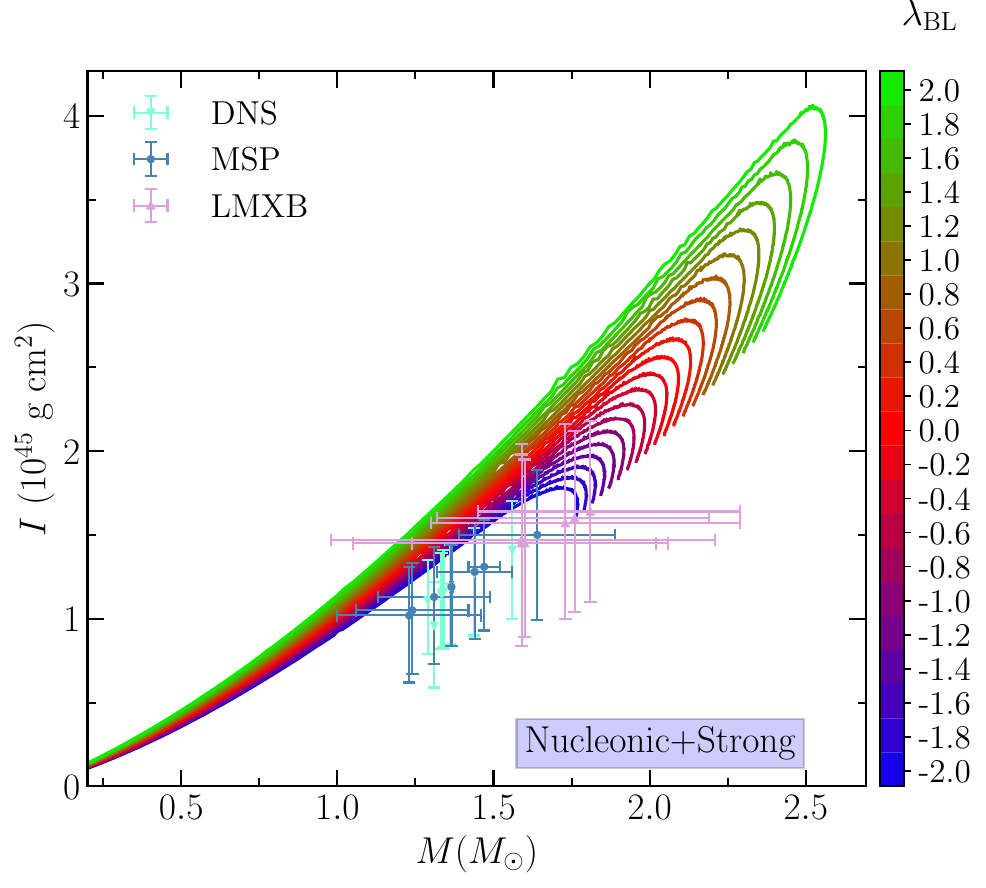} &
\includegraphics[scale=.48, angle=0]{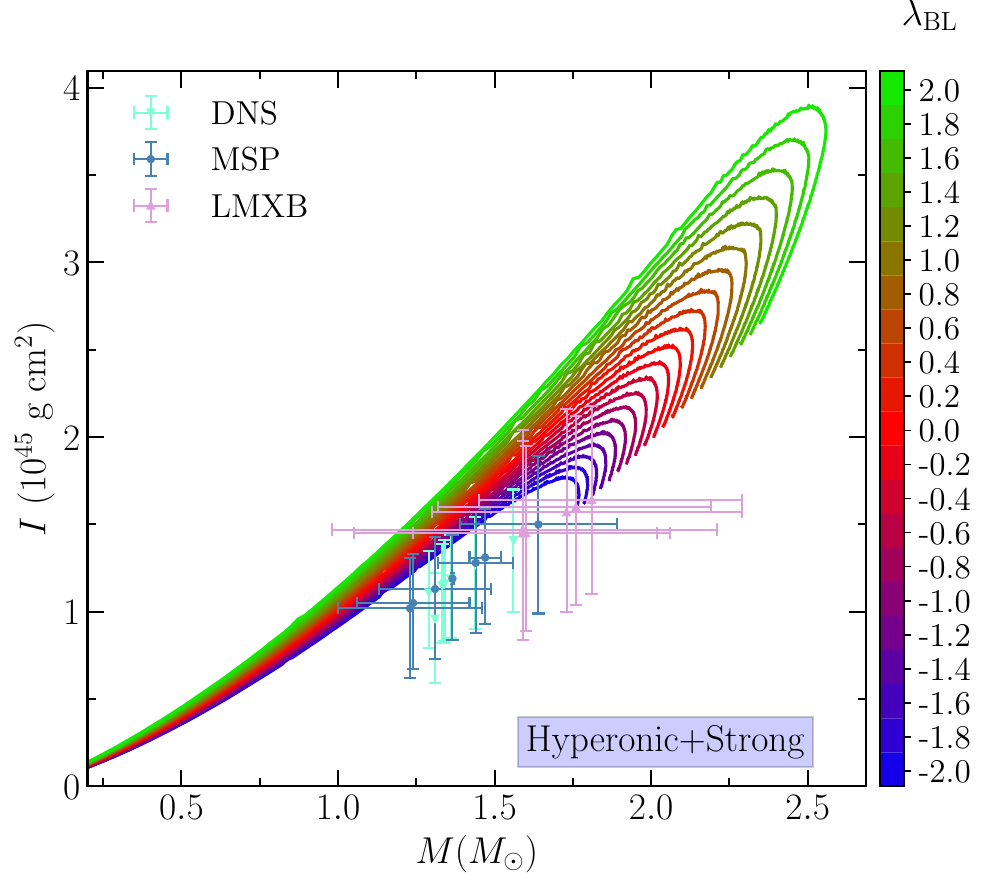}\\
\end{tabular}
\caption{Moment of inertia for different configurations of hybrid stars. The anisotropic parameter lies $-2.0~<~\lambda_{\rm BL}~<~+2.0$.}
\label{FL5}
\end{figure*}
The metric of the slowly rotating hybrid stars can be obtained within the Hartle-Throne approximation, as given in refs.~\cite{Hartle_1967, Hartle_1968, Hartle_1973}:
\begin{align}
    ds^2  = & -e^{2\nu} \ dt^2 + e^{2\lambda} \ dr + r^2 \ (d\theta^2 +\sin^2\theta d\phi^2)
    \\ \nonumber
    & - 2\omega(r)r^2\sin^2\theta \ dt \ d\phi
\end{align}
For anisotropic hybrid stars, the MOI can be calculated with the methodology developed in ref.~\cite{Rahmansyah_2020}:
\begin{align}
    I = \frac{8\pi}{3}\int_0^R \frac{r^5J\tilde{\omega}}{r- 2M}(\rho+p_r)\left[1+\frac{\sigma}{\rho+p_r}\right] \, dr,
    \label{eq:MI}
\end{align}
where $J$ = $e^{-\nu}(1-2m/r)^{1/2}$, $\tilde{\omega}=\Bar{\omega}/\Omega$ { (where $\Omega$, is the angular velocity and  $\omega$ is s the angular velocity acquired by a free-falling observer due to the drag of the inertial frame. Therefore, the angular velocity of the star relative to the local inertial frame is their difference $\bar{\omega} = \Omega -\omega$~\cite{Hartle_1973,Lopes2024ApJ}}. The final expression (using the Eq. \ref{Anisotropy_eos}) for the MOI for the anisotropic hybrid star is in the following:
\begin{align}
    I &= \frac{8\pi}{3}\int_0^R \frac{r^5J\Tilde{\omega}}{r- 2M}(\rho+p_r)\left[1+\frac{\frac{\lambda_{\rm BL}}{3} \frac{(\rho+3p_r)(\rho + p_r)r^2}{1-2m/r}}{\rho+p_r}\right] \, dr \, .
\end{align}
The MOI for four scenarios is shown in Fig. \ref{FL5} for $-2 < \lambda_{\rm BL} < +2$ values. The behavior of the MOI as the function of the mass for all cases is approximately the same; the only difference can be seen in their magnitudes, as they directly depend on the mass and radius of the star. The deduced error bars shown in the figures are taken from the ref. \cite{Kumar_2019}, predicted from the universal relations for different types of neutron stars. Except for a few curves, all lines pass through the error bars. It is observed that the higher positive values of $\lambda_{\rm BL}$ are unfavorable for the hybrid stars with those anisotropic configurations. For canonical stars, the MOI varies from (in terms of $\times~10^{45}$ g.cm$^2$) 1.44 for $\lambda_{\rm BL}$ = -2.0 up to 1.82 for $\lambda_{\rm BL}$ = +2.0. These values are fully independent of the presence of hyperons as well as the strength of the quark-quark interaction. 

However, for the maximally massive hybrid stars, we have the opposite trend. The results depend on both, the presence of hyperons or not, as well the quark-quark interaction, although the strength of the quark-quark interaction is more relevant. For instance, for $\lambda_{\rm BL}$ = -2.0, in the weak quark-quark interaction, the MOI is (in terms of $\times~10^{45}$ g.cm$^2$) 1.60 and 1.61 for Hyperonic and Nucleonic stars respectively. The same behavior is present for the strong quark-quark interaction, and the MOI is now 1.67 and 1.68 for Hyperonic and Nucleonic stars, respectively. For the isotropic case, we have the MOI equal to 2.31 (Hyperonic) and 2.32 (Nucleonic) in the weak quark-quark interaction and 2.41 (Hyperonic) and 2.44 (Nucleonic) for the strong one. Finally, for $\lambda_{\rm BL}$ = +2.0, the values are 3.64 (Hyperonic+Weak), 3.65 (Nucleonic+Weak), 3.75 (Hyperonic+Strong), and 3.87 (Nucleonic+Strong).
\section{Conclusions}
In this work, we study the influence of anisotropies on the properties of hybrid stars. We use two configurations for the hadronic EOS (with/without hyperons), as well as two strengths for the quark-quark interaction. The anisotropy is taken into account via the so-called BL model. Our main results can be summarized as:
\begin{itemize}
    \item The presence of hyperons pushes the hadron-quark phase transition to higher densities. In the same sense, increasing the quark-quark interaction ($G_V$) also makes the phase transition more difficult.

    \item { For the model used in this work}, the maximum mass is independent of the presence of hyperons. Depending only on the value of $G_V$. This is true for the isotropic case~\cite{Lopes_ApJ}, as well as for each fixed value of $\lambda_{\rm BL}$, { although it can be a model-dependent feature.}

    \item Positive values of $\lambda_{\rm BL}$ increase the maximum mass but reduce the central density. Negative values of $\lambda_{\rm BL}$ play the opposite role. As a consequence,  the minimum neutron star mass that presents deconfined quark in the core increases with $\lambda_{\rm BL}$, but the percentual amount of quark mass in the neutron stars' core decreases with $\lambda_{\rm BL}$.

    \item For a fixed $\lambda_{\rm BL}$, the value of $R_{1.4}$ is independent of the presence of hyperons and the strength of the quark-quark interaction.

    \item The constraint related to the radius of the canonical star~\cite{Miller_2021} can be satisfied assuming  -2.0 $\leq~\lambda_{\rm BL}~\leq$+0.6. About the PSR J0740+6620, we must have -0.2 $\leq~\lambda_{\rm BL}~\leq$+1.2. The range -0.2 $\leq~\lambda_{\rm BL}~\leq$+0.6 satisfies both constraints simultaneously. 

    \item The constraints related to the speculative mass of the black widow pulsars PSR J0952-0607~\cite{Romani_2022}, as well the mass-gap object in the GW190814 event~\cite{RAbbott_2020} can also be fulfilled for higher values of, $\lambda_{\rm BL}$.
    The constraint related to the HESS object~\cite{Doroshenko_2022} cannot be satisfied within the model employed in this work.

    \item The dimensionless tidal parameter $\Lambda$ decreases with $\lambda_{\rm BL}$, indicating an unexpected behavior, as it predicts that models within larger radii have lower values of $\Lambda$. All positive values of $\lambda_{\rm BL}$ are in agreement with the GW171817 constraint. Negative values of  $\lambda_{\rm BL}$ have unphysical solutions (see ref.~\cite{Biswas_2019}).

    \item The MOI increases with $\lambda_{\rm BL}$. Moreover, although most of the physical quantities related to the maximum mass are independent of the presence of hyperons, this is not true for the MOI. Nucleonic EOS presents higher values of MOI. Furthermore, inferences about the MOI from universal relations disfavor higher values of $\lambda_{\rm BL}$. Future measurements of MOI can put constraints on the degree of anisotropy.
\end{itemize}
\textbf{Acknowledgements:}

L.L.L. was partially supported by CNPq Universal Grant No. 409029/2021-1. 

\appendix

\counterwithin{figure}{section}
\section{Microscopic properties of anisotropic hybrid stars}

In the main text, we focus on the macroscopic properties of the anisotropic hybrid stars. Here we extend this subject and discuss some microscopic properties. As shown in Figs.~\ref{FL2} and Fig.~\ref{Fmq}, the main effect of a strong quark-quark interaction is to delay the hadron-quark phase transition, pushing it to higher densities and higher star masses. Therefore, here we only discuss the weak quark-quark phase transition with and without hyperons. The qualitative discussion for a strong one is the same.

\begin{figure*}[tb]
\begin{tabular}{ccc}
\centering 
\includegraphics[scale=.48, angle=0]{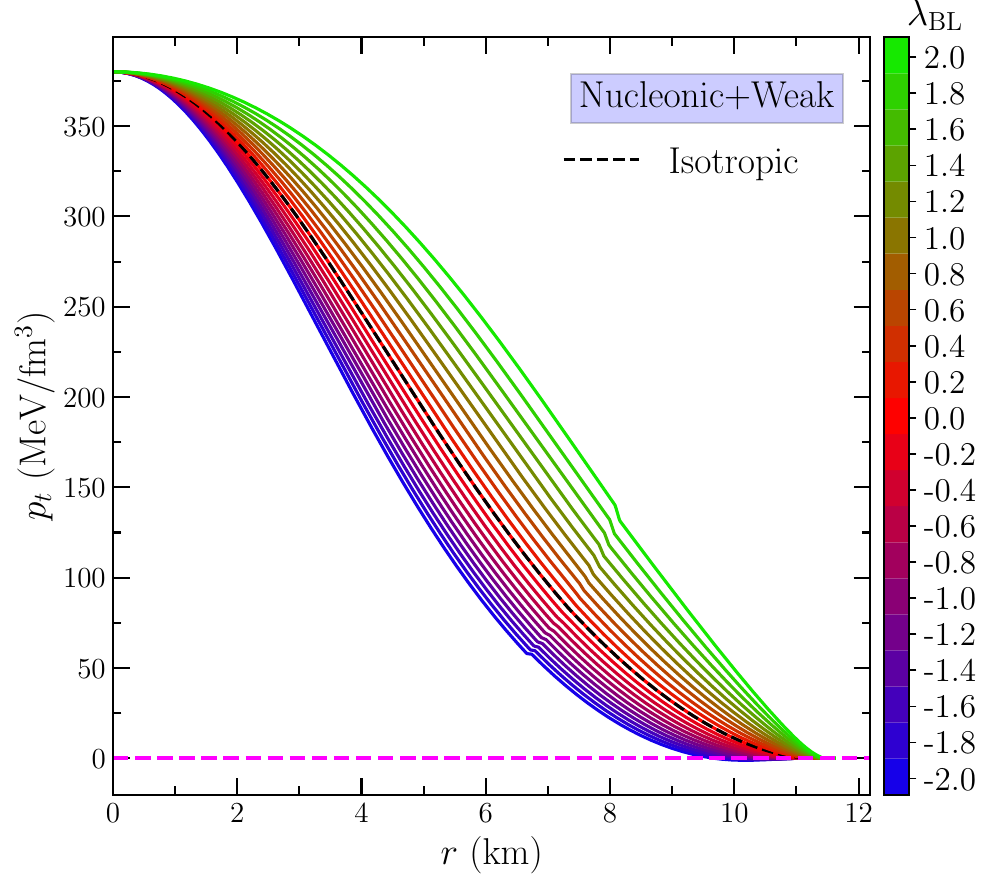} &
\includegraphics[scale=.48, angle=0]{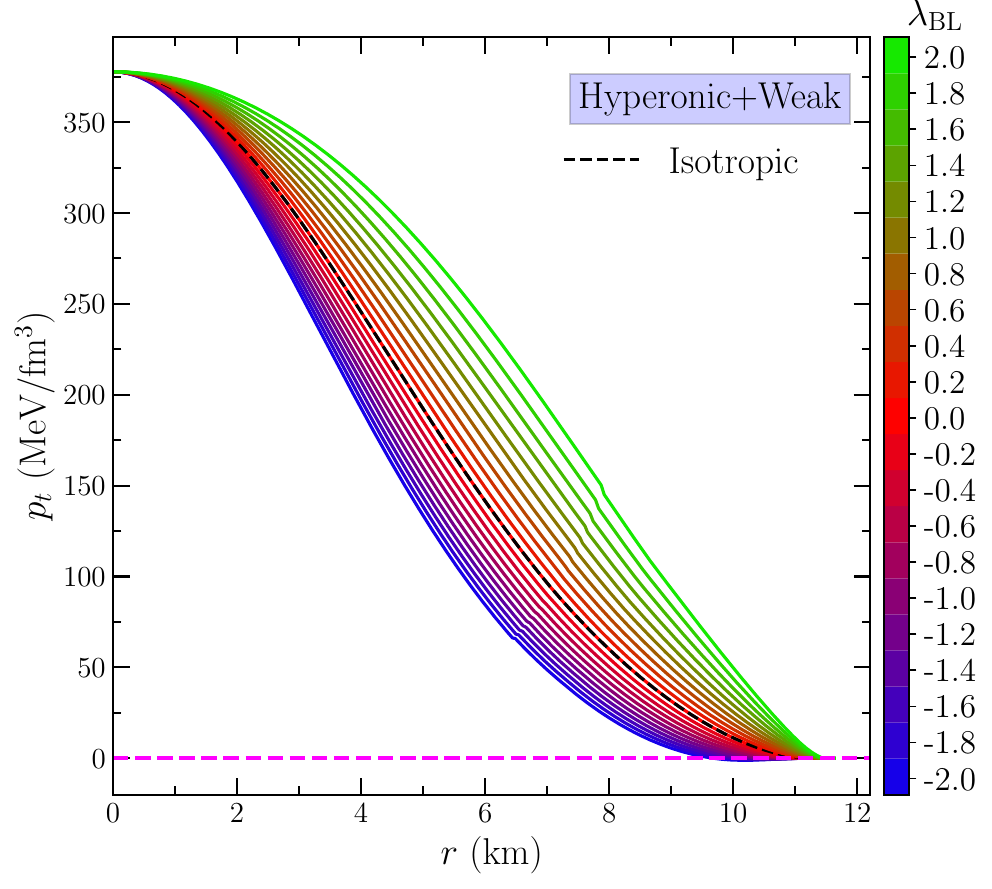}\\
\end{tabular}
\caption{Both tangential and radial pressure profiles for maximum mass configurations are shown.}
\label{App:FigA1}
\end{figure*}

\begin{figure*}[tb]
\begin{tabular}{ccc}
\centering 
\includegraphics[scale=.48, angle=0]{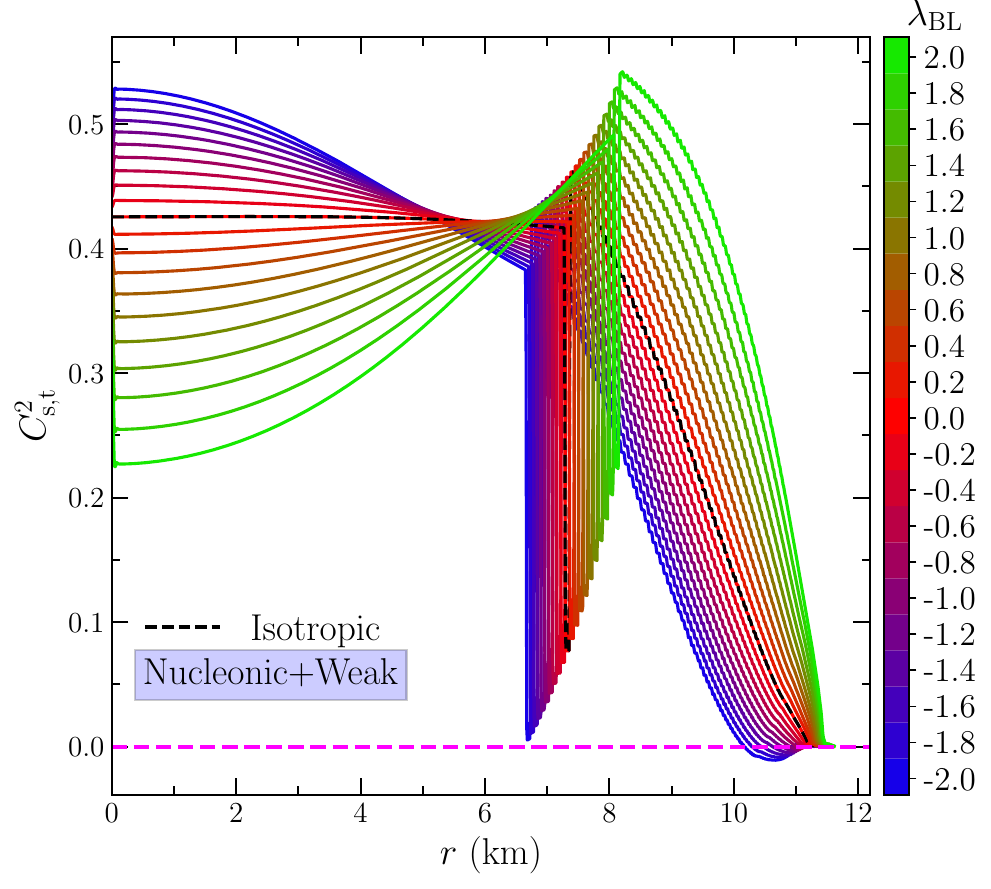} &
\includegraphics[scale=.48, angle=0]{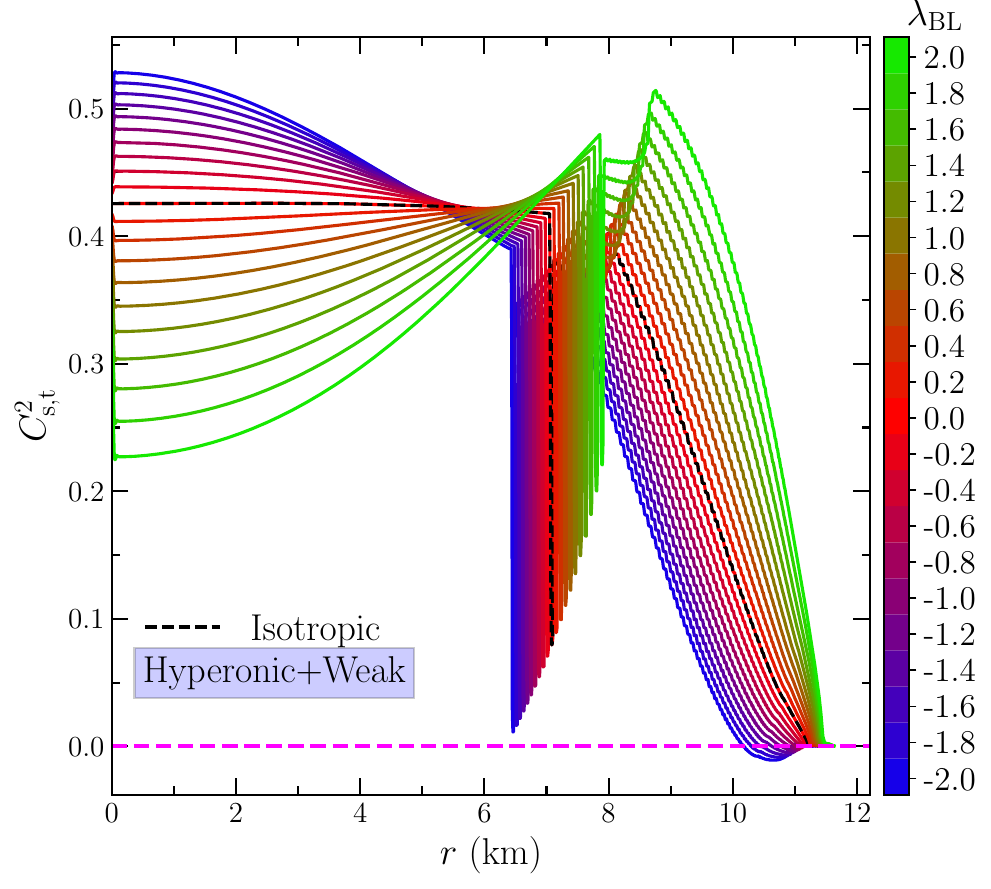}\\
\end{tabular}
\caption{Speed of sound profiles for maximum mass configurations are shown.}
\label{App:FigA2}
\end{figure*}

Using the maximally massive star, we begin studying how the tangential pressure behaves in function of the star radius. The numerical results are presented in Fig.~\ref{App:FigA1}. As we can see, for most of the star's interior, the pressure kept its monotonic behavior and decreased with the radius. However, near the star's surface, for negative values of $\lambda_{BL}$, the tangential pressure becomes slightly negative, once for low values of $\rho$, the second term in Eq.~\ref{Anisotropy_eos} dominates over the first one, as $\rho >> p_r$. The radial pressure is the same of the isotropic pressure, displayed as a dotted black line.

As we approach the stellar surface, the energy density and the radial pressure go to zero, making the tangential pressure zero as well. For negative values of $\lambda_{BL}$, this implies the existence of a region where the tangential pressure increases with the radius, which produces negative values for the speed of sound. This can be easily be seen in Fig.~\ref{App:FigA2}.

Starting from the isotropic case, we can also notice from Fig.~\ref{App:FigA2} that the speed of sound is almost constant in the core (quark phase), presenting a discontinuity at the hadron-quark phase transition and then decreases with the radius in the hadron phase. In the presence of anisotropy, we have different behaviors at the core. For positive values of $\lambda_{BL}$ the speed of sound increases up to the hadron-quark phase transition and then decreases. For negative values, the speed of sound is essentially decreased with the radius (plus the discontinuity at the hadron-quark phase transition). However, as pointed out earlier, close to the surface it becomes negative. It is this region that produces the  unphysical results for the dimensionless tidal parameter and the reason why we restrict Fig.~\ref{FL4} in the range $0~<~\lambda_{BL}~<2.0$.

\newpage

\bibliography{ASQS}
\end{document}